\documentclass[final,5p,times,number]{elsarticle}
\usepackage{graphicx}
\usepackage{subcaption}     
\usepackage{lscape}
\usepackage{multirow}
\usepackage{booktabs}
\usepackage{float}
\usepackage{siunitx}
\biboptions{sort&compress}
\usepackage{amsmath}
\usepackage{amssymb}
\usepackage{ifthen,multicol}
\usepackage[noprefix]{nomencl}
\usepackage{ifpdf}\ifpdf 
\usepackage[colorlinks,linktocpage,pdftex,hyperfigures]{hyperref}
\else
\providecommand{\href}[2]{#2}
\providecommand{\url}[1]{}
\providecommand{\URLprefix}{}
\fi

\renewcommand{\nomgroup}[1]{\medskip
\ifthenelse{\equal{#1}{B}}{\item[\textit{Abbreviations}]}{ \ifthenelse{\equal{#1}{G}}{\item[\textit{Greek
symbols}]}{ \ifthenelse{\equal{#1}{S}}{\item[\textit{Subscripts}]}{
\ifthenelse{\equal{#1}{P}}{\item[\textit{Superscripts}]}{}}}}}
 \makenomenclature
\journal{Int. J. Therm. Sci.}

\begin{document}

\begin{frontmatter}

\title{Experimental analysis and transient numerical simulation of a large diameter pulsating heat pipe in microgravity conditions}

\author[1]{Mauro Abela}
\author[1]{Mauro Mameli}
\author[2]{Vadim Nikolayev}
\author[1]{Sauro Filippeschi}
\address[1]{Department of Energy, Systems Land and Construction Engineering, University of Pisa, Largo L. Lazzarino, Pisa, Italy}
\address[2]{Service de Physique de l'Etat Condens\'e, CEA, CNRS, Universit\'e Paris-Saclay, CEA Saclay, 91191 Gif-sur-Yvette Cedex, France}

\begin{abstract}
A multi-parametric transient numerical simulation of the start-up of a large diameter Pulsating Heat Pipe (PHP) specially designed for future experiments on the International Space Station (ISS) are compared to the results obtained during a parabolic flight campaign supported by the European Space Agency. Since the channel diameter is larger than the capillary limit in normal gravity, such a device behaves as a loop thermosyphon on ground and as a PHP in weightless conditions; therefore, the microgravity environment is mandatory for pulsating mode. Because of a short duration of microgravity during a parabolic flight, the data concerns only the transient start-up behavior of the device. One of the most comprehensive models in the literature, namely the in-house 1-D transient code CASCO (French acronym for Code Avanc\'e de Simulation du Caloduc Oscillant: Advanced PHP Simulation Code in English), has been configured in terms of geometry, topology, material properties and thermal boundary conditions to model the experimental device.
The comparison between numerical and experimental results is performed simultaneously on the temporal evolution of multiple parameters: tube wall temperature, pressure and, wherever possible, velocity of liquid plugs, their length and temperature distribution within them. The simulation results agree with the experiment for different input powers. Temperatures are predicted with a maximum deviation of 7\%. Pressure variation trend is qualitatively captured as well as the liquid plug velocity, length and temperature distribution. The model also shows the ability of capturing the instant when the fluid pressure begins to oscillate after the heat load is supplied, which is a fundamental information for the correct design of the engineering model that will be tested on the ISS.  We also reveal the existence of strong liquid temperature gradients near the ends of liquid plugs both experimentally and by simulation. Finally, a theoretical prediction of the stable functioning of a large diameter PHP in microgravity is given. Results show that the system provided with an input power of \SI{185}{W} should be able to reach the steady state after \SI{1}{min} and maintain a stable operation from then on.
\end{abstract}


\end{frontmatter}
\makeatletter
\twocolumn[%
\begin{@twocolumnfalse}
\fbox{ 
\begin{minipage}{\textwidth}
\begin{multicols}{2}

\begin{thenomenclature}
\nomgroup{A}
  \item [{$\cal L$}]\begingroup latent heat [\si{J/kg}]\nomeqref {0}\nompageref{2}
  \item [{$C$}]\begingroup thermal mass [\si{J/K}]\nomeqref {0}\nompageref{2}
  \item [{$c$}]\begingroup specific heat [\si{J/(kg.K)}]\nomeqref {0}\nompageref{2}
  \item [{$D$}]\begingroup heat diffusivity [\si{m^2/s}]\nomeqref {0}\nompageref{2}
  \item [{$F$}]\begingroup viscous friction force [\si{N}]\nomeqref {0}\nompageref{2}
  \item [{$f$}]\begingroup acquisition frequency [\si{Hz}]\nomeqref {0}\nompageref{2}
  \item [{$g$}]\begingroup effective gravity acceleration [\si{m^2/s}]\nomeqref {0}\nompageref{2}
  \item [{$j$}]\begingroup volume heat generation rate [\si{W/m^3}]\nomeqref {0}\nompageref{2}
  \item [{$L$}]\begingroup length [\si{m}]\nomeqref {0}\nompageref{2}
  \item [{$M$}]\begingroup total number of bubbles or plugs\nomeqref {0}\nompageref{2}
  \item [{$m$}]\begingroup mass of vapor [\si{kg}]\nomeqref {0}\nompageref{2}
  \item [{$N$}]\begingroup total number\nomeqref {0}\nompageref{2}
  \item [{$Nu$}]\begingroup Nusselt number\nomeqref {0}\nompageref{2}
  \item [{$P$}]\begingroup power [\si{W}]\nomeqref {0}\nompageref{2}
  \item [{$p$}]\begingroup pressure; $p_i$: in the bubble $i$ [\si{Pa}]\nomeqref {0}\nompageref{2}
  \item [{$q$}]\begingroup heat flux [\si{W/m^2}]\nomeqref {0}\nompageref{2}
  \item [{$r$}]\begingroup tube inner radius [\si{m}]\nomeqref {0}\nompageref{2}
  \item [{$R_v$}]\begingroup vapor gas constant [\si{J/(kg.K)}]\nomeqref {0}\nompageref{2}
  \item [{$Re$}]\begingroup liquid Reynolds number\nomeqref {0}\nompageref{2}
  \item [{$S$}]\begingroup cross-section area [\si{m^2}]\nomeqref {0}\nompageref{2}
  \item [{$T$}]\begingroup temperature ($T_i$: of vapor) [\si{K}]\nomeqref {0}\nompageref{2}
  \item [{$t$}]\begingroup time [\si{s}]\nomeqref {0}\nompageref{2}
  \item [{$U$}]\begingroup heat transfer coefficient, conductance [\si{W/(m^2.K)}]\nomeqref {0}\nompageref{2}
  \item [{$V$}]\begingroup liquid velocity [\si{m/s}]\nomeqref {0}\nompageref{2}
  \item [{$x,X$}]\begingroup abscissa measured along the PHP tube [\si{m}]\nomeqref {0}\nompageref{2}
\nomgroup{B}
  \item [{ANN}]\begingroup artificial neural networks\nomeqref {0}\nompageref{2}
  \item [{CASCO}]\begingroup Advanced PHP simulation code (in French)\nomeqref {0}\nompageref{2}
  \item [{CFD}]\begingroup computational fluid dynamics\nomeqref {0}\nompageref{2}
  \item [{EOS}]\begingroup equation of state\nomeqref {0}\nompageref{2}
  \item [{FEC}]\begingroup film evaporation/condensation\nomeqref {0}\nompageref{2}
  \item [{IR}]\begingroup infra red\nomeqref {0}\nompageref{2}
  \item [{ISS}]\begingroup International Space Station\nomeqref {0}\nompageref{2}
  \item [{PHP}]\begingroup pulsating heat pipe\nomeqref {0}\nompageref{2}
  \item [{SMD}]\begingroup spring-mass-damper\nomeqref {0}\nompageref{2}
\nomgroup{G}
  \item [{$\delta$}]\begingroup thickness [\si{m}]\nomeqref {0}\nompageref{2}
  \item [{$\gamma$}]\begingroup vapor adiabatic index$=c_{v,p}/c_{v,v}$\nomeqref {0}\nompageref{2}
  \item [{$\lambda$}]\begingroup heat conductivity [\si{W/(m.K)}]\nomeqref {0}\nompageref{2}
  \item [{$\nu$}]\begingroup liquid kinematic viscosity [\si{m^2/s}]\nomeqref {0}\nompageref{2}
  \item [{$\Omega$}]\begingroup vapor bubble volume [\si{m^3}]\nomeqref {0}\nompageref{2}
  \item [{$\rho$}]\begingroup density [\si{kg/m^3}]\nomeqref {0}\nompageref{2}
\nomgroup{P}
  \item [{$exp$}]\begingroup experimental\nomeqref {0}\nompageref{2}
  \item [{$l$}]\begingroup left\nomeqref {0}\nompageref{2}
  \item [{$r$}]\begingroup right\nomeqref {0}\nompageref{2}
  \item [{$s$}]\begingroup $r$ or $l$\nomeqref {0}\nompageref{2}
  \item [{$sens$}]\begingroup sensible\nomeqref {0}\nompageref{2}
  \item [{$sim$}]\begingroup simulation\nomeqref {0}\nompageref{2}
  \item [{$thr$}]\begingroup threshold\nomeqref {0}\nompageref{2}
\nomgroup{S}
  \item [{$a$}]\begingroup adiabatic\nomeqref {0}\nompageref{2}
  \item [{$c$}]\begingroup condenser\nomeqref {0}\nompageref{2}
  \item [{$cons$}]\begingroup mass conservation\nomeqref {0}\nompageref{2}
  \item [{$e$}]\begingroup evaporator, external\nomeqref {0}\nompageref{2}
  \item [{$f$}]\begingroup liquid film\nomeqref {0}\nompageref{2}
  \item [{$fb$}]\begingroup feedback section (vertical in Fig.~\ref{Init})\nomeqref {0}\nompageref{2}
  \item [{$i$}]\begingroup bubble or plug identifier\nomeqref {0}\nompageref{2}
  \item [{$l$}]\begingroup liquid\nomeqref {0}\nompageref{2}
  \item [{$m$}]\begingroup meniscus\nomeqref {0}\nompageref{2}
  \item [{$next$}]\begingroup next to the bubble $i$\nomeqref {0}\nompageref{2}
  \item [{$nucl$}]\begingroup nucleated\nomeqref {0}\nompageref{2}
  \item [{$p$}]\begingroup PHP spatial period or at constant pressure\nomeqref {0}\nompageref{2}
  \item [{$s$}]\begingroup spreader\nomeqref {0}\nompageref{2}
  \item [{$sat$}]\begingroup at saturation\nomeqref {0}\nompageref{2}
  \item [{$t$}]\begingroup total\nomeqref {0}\nompageref{2}
  \item [{$v$}]\begingroup vapor or at constant volume\nomeqref {0}\nompageref{2}
  \item [{$w$}]\begingroup internal tube wall or its material\nomeqref {0}\nompageref{2}

\end{thenomenclature}

\end{multicols}
\end{minipage} }\vspace*{0.5truecm}\end{@twocolumnfalse} ] \makeatother

\nomenclature[p]{$sens$}{sensible}
\nomenclature[p]{$exp$}{experimental}
\nomenclature[p]{$sim$}{simulation}
\nomenclature[p]{$l$}{left}
\nomenclature[p]{$r$}{right}
\nomenclature[p]{$s$}{$r$ or $l$}
\nomenclature[s]{$m$}{meniscus}
\nomenclature[s]{$e$}{evaporator, external}
\nomenclature[s]{$s$}{spreader}
\nomenclature[s]{$f$}{liquid film}
\nomenclature[s]{$p$}{PHP spatial period or at constant pressure}
\nomenclature[s]{$l$}{liquid}
\nomenclature[s]{$a$}{adiabatic}
\nomenclature[s]{$t$}{total}
\nomenclature[s]{$c$}{condenser}
\nomenclature[s]{$v$}{vapor or at constant volume}
\nomenclature[s]{$w$}{internal tube wall or its material}
\nomenclature[s]{$i$}{bubble or plug identifier}
\nomenclature[s]{$next$}{next to the bubble $i$}
\nomenclature[s]{$cons$}{mass conservation}
\nomenclature[s]{$sat$}{at saturation}
\nomenclature[s]{$nucl$}{nucleated}
\nomenclature[s]{$fb$}{feedback section (vertical in Fig.~\ref{Init})}
\nomenclature[p]{$thr$}{threshold}
\nomenclature[a]{$\cal L$}{latent heat [\si{J/kg}]}
\nomenclature[a]{$f$}{acquisition frequency [\si{Hz}]}
\nomenclature[a]{$Nu$}{Nusselt number}
\nomenclature[a]{$P$}{power [\si{W}]}
\nomenclature[a]{$q$}{heat flux [\si{W/m^2}]}
\nomenclature[a]{$j$}{volume heat generation rate [\si{W/m^3}]}
\nomenclature[a]{$D$}{heat diffusivity [\si{m^2/s}]}
\nomenclature[a]{$M$}{total number of bubbles or plugs}
\nomenclature[a]{$N$}{total number}
\nomenclature[a]{$T$}{temperature ($T_i$: of vapor) [\si{K}]}
\nomenclature[a]{$x,X$}{abscissa measured along the PHP tube [\si{m}]}
\nomenclature[a]{$g$}{effective gravity acceleration [\si{m^2/s}]}
\nomenclature[a]{$L$}{length [\si{m}]}
\nomenclature[a]{$m$}{mass of vapor [\si{kg}]}
\nomenclature[a]{$V$}{liquid velocity [\si{m/s}]}
\nomenclature[a]{$c$}{specific heat [\si{J/(kg.K)}]}
\nomenclature[a]{$C$}{thermal mass [\si{J/K}]}
\nomenclature[a]{$U$}{heat transfer coefficient, conductance [\si{W/(m^2.K)}]}
\nomenclature[a]{$F$}{viscous friction force [\si{N}]}
\nomenclature[a]{$p$}{pressure; $p_i$: in the bubble $i$ [\si{Pa}]}
\nomenclature[a]{$t$}{time [\si{s}]}
\nomenclature[a]{$Re$}{liquid Reynolds number}
\nomenclature[a]{$r$}{tube inner radius [\si{m}]}
\nomenclature[a]{$S$}{cross-section area [\si{m^2}]}
\nomenclature[a]{$R_v$}{vapor gas constant [\si{J/(kg.K)}]}
\nomenclature[g]{$\rho$}{density [\si{kg/m^3}]}
\nomenclature[g]{$\nu$}{liquid kinematic viscosity [\si{m^2/s}]}
\nomenclature[g]{$\gamma$}{vapor adiabatic index$=c_{v,p}/c_{v,v}$}
\nomenclature[g]{$\delta$}{thickness [\si{m}]}
\nomenclature[g]{$\lambda$}{heat conductivity [\si{W/(m.K)}]}
\nomenclature[g]{$\Omega$}{vapor bubble volume [\si{m^3}]}
\nomenclature[b]{EOS}{equation of state}
\nomenclature[b]{CASCO}{Advanced PHP simulation code (in French)}
\nomenclature[b]{IR}{infra red}
\nomenclature[b]{PHP}{pulsating heat pipe}
\nomenclature[b]{SMD}{spring-mass-damper}
\nomenclature[b]{CFD}{computational fluid dynamics}
\nomenclature[b]{ANN}{artificial neural networks}
\nomenclature[b]{FEC}{film evaporation/condensation}
\nomenclature[b]{ISS}{International Space Station}

\section{Introduction}
In recent years, the market growing interest in high performance, high reliability and low cost heat transfer devices, drew research attention to an innovative technology: the Pulsating Heat Pipe (PHP). Invented in early 90's by \citet{akachi}, the PHP is a simple capillary tube that meanders between a heat source and a cooler (where the tube sections become evaporators and condensers, respectively). Due to capillary forces, the working fluid resides inside the tube as an alternation of vapor bubbles and liquid plugs which oscillate during device operation. Technological simplicity (and therefore, reliability) and high heat transfer performance widen its possible field of applications from the thermal management of electronic devices to space subsystems thermal control where high reliability and passive operation play a key role.

The interest to the adoption of PHP for space applications is witnessed by the growing number of investigations of PHP performance under weightlessness conditions.  \citet{Gu05,Ayel15,Taft15,Paiva10,Paiva14} carried on experiments during parabolic flights, while  \citet{Daimaru17a,Ando18,Taft19} conducted long term on-orbit experiments thus making the PHP technology reach a high TRL.

Moreover, space environment offers a favorable condition for the enhancement of PHP thermal performance. In fact, in microgravity the ratio between buoyancy forces and surface tension decreases; in this way it is possible to enhance the heat transfer capability by increasing the pipe diameter beyond the capillary limit on ground. This possibility was first speculated by \citet{Gu05} and later proved on board a parabolic flight with a tubular PHP filled with FC-72 by \citet{Mangini17} and  \citet{Mameli19}. \citet{Cecere18} and  \citet{Ayel19} tested a copper flat PHP filled respectively with a self-rewetting fluid and FC-72 during the parabolic flights. However, in all the above cases, due to the short duration of microgravity periods, it was not possible to reach a pseudo-steady state. For this reason, long term tests in a microgravity environment are mandatory for a complete characterization of thermal performance of a large diameter PHP. To do so, ESA promoted the development of a large diameter PHP that will be implemented on the Heat Transfer Host of the ISS where it will undergo long time test in microgravity environment.

Despite the experience gained in PHP field \cite{Kim21c}, simulation tools are needed to optimize the design of such devices.  Since no steady-state data is available for large diameter PHP, it is inevitable to validate models on the start-up transient behavior. Unfortunately, only few models present in literature are able to perform a transient simulation and even fewer are validated against experimental data.  This is primarily due to the intricate weave of co-acting mechanisms governing the PHP operation.
For this reason, since its introduction, many researchers are trying to thoroughly model the PHP by using different approaches for the description of the primary operational mode (i.e. the plug-slug flow). These attempts can be classified as follows: (1) Continuum wave propagation approach, in which pressure oscillations are fundamental to induce vapor-liquid circulation; (2) Spring-mass-damper (SMD) approach, in which the PHP is modeled as a single or multiple SMD system (liquid plugs are modeled as masses, vapor bubbles as non-linear springs, friction and the capillary forces as non-linear dampers); (3) Artificial Neural Networks (ANN), a statistical data modeling inspired by learning processes of human brain; (4) Empirical correlations based on dimensionless groups; (5) 2D and 3D approaches using freeware CFD tools or commercial software; (6) 1D approach based on a set of averaged equations of hydrodynamics with phase change (mass, momentum, energy, etc.); this is largely the most adopted approach. Here we give only a brief account of the existing literature; a much more detailed review can be found in the work \cite{ATE21}.

First, the models \cite{Miyazaki96,miyazaki,Zuo01} belonging to the classes (1) and (2) have appeared. Generally, they attempted to describe a particular PHP functioning regime with a phenomenological description. As a result, the application of such models is limited to a specific regime. \citet{Sun17a} used an SMD model to investigate the effects of filling ratio, tube length, inner diameter, temperature difference between the evaporator and condenser sections, and working fluid on oscillating motion characteristics of liquid plugs and bubbles of a PHP in micro-gravity condition; however the model is not validated against experimental data. More recently, \citet{Yoon19} developed an SMD model to theoretically analyze the liquid plugs oscillation dominant frequency observed experimentally in the small-amplitude oscillation regime.

Contemporary growth of interest in ANN, as result of the growing computing capabilities, has pushed the development of models belonging to the class (3). The first attempt of this kind is the one of \citet{Khandekar02} who trained an ANN with 52~sets of experimental data in order to predict the equivalent thermal resistance. Similarly, \citet{Patel18} discussed a performance prediction model based on an ANN trained with 1652 copper PHPs data collected from literature between 2003 and 2017 which was able to predict the thermal resistance with high prediction accuracy (with a coefficient of determination $R^2=0.89$). \citet{Wang19a} used the same approach obtaining similar results. It is worth to notice that in the discussed cases ANN models are used to predict the overall performance (i.e. overall thermal resistance) rather than the evolution of specific parameters (i.e. temperatures and pressures of a particular spot); this is a consequence of the nature of ANN approach which is limited by the training dataset and acts as a black-box hiding physical bases of the phenomena involved in the PHP functioning; empirical correlation approaches (4) are affected by the same limitations.
As an example, \citet{Shafii10} used their experimental data along with the data collected from literature to develop a power-law correlation for the input heat flux prediction; their results show a good agreement with experimental data (88.6\% of the deviations are within $\pm 30$\%). A similar approach is also presented in the above cited work \cite{Patel18} where a linear and a power-law correlations agree with those obtained by \citeauthor{Shafii10}. However, because of multitude of the relevant physical parameters that impact the PHP functioning, the approaches belonging to the classes (3--4) can hardly be considered as candidates for a design tool of a PHP with an arbitrary structure.

Unlike the above approaches, those belonging to the classes (5--6) try to provide a comprehensive physical description of fundamental phenomena involved in the PHP functioning (vapor bubble nucleation and coalescence, liquid film dynamics etc.); in this way, it is possible not only to capture the overall performance in some particular regimes, but also describe transitions between them and consider the temporal evolution of various PHP parameters; the latter feature is essential for such a non-stationary system as the PHP. \citet{Vo20} modeled the 3D flow in PHP by using the ANSYS Fluent\textsuperscript{\textregistered} software that captured qualitatively the circulating plug motion observed experimentally. With the same software, \citet{Wang20} simulated a miniature single loop PHP in 2D to investigate the effects of tube constrictions in the condenser zone. The model is able to reproduce some key phenomena observed experimentally such as nucleate boiling, formation of liquid plugs, coalescence of vapor bubbles and flow patterns transition; moreover, validation is performed on steady state evaporator average temperature and overall thermal resistance. However, because of the difficulties to describe adequately the free vapor-liquid interfaces with phase change and huge computational costs, the multidimensional modeling can hardly be considered as a viable tool.

As of now, the 1D approaches belonging to the class (6) represent the best compromise between reasonable computational costs and a thorough physical description. For this reason, it is the most suitable way to provide a simulation tool for the design and study of PHP prototypes. \citet{Wong99} proposed the first model of this kind, which is based on the solution of a set of first order non-linear differential equations to describe an adiabatic flow in a capillary channel; however, this model and its sequels appeared in the beginning of 2000's still neglected many relevant mechanisms (in particular, the liquid film evaporation among others). \citet{Holley05} proposed the first comprehensive model able to account for liquid plugs coalescence and bubble nucleation; this was later improved by \citet{Mameli12,MameliMST12,Manzoni16,Manzoni16a} implementing tube bends effects and two-phase heat transfer coefficient calculation as function of the heating regimes. \citet{shafii1} proposed the equations for the bubble and plug dynamics with a solid hydrodynamic and thermodynamic background.  Starting from this platform, \citet{IJHMT10} developed the film evaporation/condensation (FEC) model for a single liquid/vapor couple able to explain the large amplitude oscillations observed experimentally. \citet{JHT11} extended the FEC model to treat an arbitrary number of bubbles and branches which takes into account many important phenomena, such as coalescence of liquid plugs and film dynamics. In the same work, the basic architecture of the in-house C++ software called CASCO (Code Avanc\'e de Simulation du Caloduc Oscillant: Advanced PHP simulation code in French) was proposed. It is based on the FEC model. CASCO was later updated \cite{IarATE17} to account for the tube heat conduction and bubble nucleation; it was later used \cite{MST19} to study the impact of orientation with respect to gravity on the PHP performance and to explain different PHP functioning regimes \cite{ISOPHP19}. However, CASCO still lacks of a thorough validation. By following the main principles of CASCO, \citet{Daimaru17} developed a simplified numerical model (where only one dry spot per vapor bubble was allowed). The model is validated it against the data of an on-orbit experiment and revealed the energy propagation as reason for the pressure propagation.

By using the film model of \citet{Senjaya14}, \citet{Bae17} have simulated the spatial and temporal variations of the liquid film thickness; the model is able to predict overall thermal performance of PHP in various orientations. \citet{Noh20} used this model to perform a numerical optimization of PHP in terms of channel diameter and number of turns and proposed a merit number that can be used as a guideline for PHP design under the constraint of fixed space.

In Table~\ref{ModelsTab}, an overview of the above mentioned models is presented. It clearly appears that a crucial step for the development of a comprehensive simulation tool is still missing. Most of the cited models are validated in steady-state conditions and on a single parameter (i.e. overall thermal resistance, average evaporator temperature etc.) while only a few authors make comparisons with the experiment of several physical parameters at once. However, phenomena having a major impact on PHP performance --- such as start-up and dryout --- are intrinsically transient and depend on the interplay of multiple parameters (local temperatures, pressures and fluid thermodynamic state). Therefore, any model that aims to be predictive needs to be validated simultaneously on multiple parameters in transient conditions so as to prove the ability to capture different phenomena occurring during PHP operation.

The user interface of the CASCO software has a capability to input any PHP structure, in particular an arbitrary number and positions of heat sources and sinks. It is able to provide the evolution of various parameters (solid wall temperature, liquid plug temperature, bubble pressure etc.) as functions of time and spatial coordinate that can be compared with the experiment. This ability along with reasonably low computational costs (\SI{\approx 5}{min}  for \SI{20}{s} simulation) makes CASCO one of the most suitable candidates as a comprehensive PHP simulation tool.

Aiming to contribute to the definition of a suitable numerical model for a complete simulation of a large diameter PHP, objective of this paper is to show the prediction ability of the FEC model by comparing simulation results with the experimental data. CASCO is used to simulate a large diameter PHP designed for a future implementation on the heat transfer host (HTH) apparatus onboard the International Space Station (ISS). Since the tube internal diameter is larger than the static capillary threshold evaluated for the working fluid (FC72) in normal gravity conditions, the device behaves as a thermosyphon and as PHP only in micro-gravity conditions. For this reason, the data used for the validation is collected in the micro-gravity environment of the  67$^{th}$ ESA parabolic flight campaign. Since the short duration of microgravity periods is insufficient to reach the steady (or rather pseudo steady) state operation, the experimental data refers to device start-up only.
The comparison is performed simultaneously on the temporal evolution of multiple parameters: temperatures, pressures and local fluid characteristics. The latter comprise liquid plugs velocities and lengths. Results suggest that the model is able to closely reproduce all the key phenomena observed experimentally. A theoretical prediction of the steady state behavior of  a large diameter PHP in weightless conditions are reported. Another objective of the paper is to study the spatial variation of temperature inside the liquid plugs and reveal the existence of strong thermal gradients near the plug ends both experimentally and by simulation.
\begin{landscape}
\begin{table}[ht]
  \begin{tabular}{lllll}
  \hline
  Approach& References                                                                                                                         & \begin{tabular}[c]{@{}l@{}} Comparison\\ with experiment\end{tabular}                           &\begin{tabular}[c]{@{}l@{}} Validation\\ parameters\end{tabular}                                                                             & Remarks  \\ \midrule
   & Miyazaki et al.  \cite{Miyazaki96,miyazaki}                                                     &  fair  & single                     &   \begin{tabular}[c]{@{}l@{}} Correctly predicts the experimental pressure\\  wave velocity.   \end{tabular}                                                                                                                             \\
                                                                  \begin{tabular}[c]{@{}l@{}} (1) Continuum \\~Wave Propagation\end{tabular} & & & &  \\
                                                                        &\citet{Zuo01} & fair  & single                                                                                        & \begin{tabular}[c]{@{}l@{}}The model predicts the proper  filling ratio.\end{tabular}          \\ \hline
                                                                         &\citet{Sun17a}                                                                                      &none &  -     &  \begin{tabular}[c]{@{}l@{}} Theoretical study of the effect of various \\parameters on oscillation characteristics in \\micro-gravity.  \end{tabular}  \\
                                         (2) SMD models                   & & & &  \\
                                                                  &\citet{Yoon19}                                                                                     &   fair  & single                                                                                                                       &\begin{tabular}[c]{@{}l@{}}Theoretical investigation of dominant frequencies\\of liquid plugs oscillation. \end{tabular}          \\ \hline
                                                                        &\citet{Khandekar02}                                      &       average     & single                                                                                                                 & \begin{tabular}[c]{@{}l@{}} An ANN is trained using a set of 52~sets of \\experimental data to predict thermal \\resistance. \end{tabular}            \\
  & & & &  \\
  (3) ANN approaches                                                                   &\citet{Patel18}                                                                                    &    average & single                                                                                                                     &\begin{tabular}[c]{@{}l@{}} ANN trained using 1652 experimental sets \\to predict thermal resistance. \end{tabular}            \\

                                                                        &\citet{Wang19a}                                                                                     &   average&	single                                                                                                                      &\begin{tabular}[c]{@{}l@{}} The model is used to directly predict the\\ thermal resistance of the PHP with various \\working fluids.  The influence of geometry,\\ property parameters and operational\\ parameters are considered by using  a \\non-dimensional group. \end{tabular}          \\ \hline
                                                    &\citet{Shafii10}                                                                                 & average & single                                             &  \begin{tabular}[c]{@{}l@{}} A power law correlation based on 	a non-dimen-\\sional group is used to predict  input heat flux.\end{tabular}        \\
	\begin{tabular}[c]{@{}l@{}} (4) Empirical\\ correlations    \end{tabular}   														& & & &  \\
                                                  &\citet{Patel18}                                                                                    &   average & single                                                                                                                      & \begin{tabular}[c]{@{}l@{}} A linear and a power law correlations are used \\to predict input heat flux.\end{tabular}            \\ \hline

                                                                         &\citet{Vo20}                                                                                          &     average &  \begin{tabular}[c]{@{}l@{}} multiple:\\circulation regime\\steady-state heat transfer rate \end{tabular}                                                                                                                  &    \begin{tabular}[c]{@{}l@{}} 3D ANSYS Fluent model\\for multi-branch PHP.  \end{tabular}              \\

							(5) CFD models  & & & &  \\
																 &\citet{Wang20}                                                                                    & good & \begin{tabular}[c]{@{}l@{}} multiple:\\steady-state evaporator temperature\\overall thermal resistance \end{tabular}                                                                                                                         & \begin{tabular}[c]{@{}l@{}} A 2D model is developed with ANSYS Fluent to study effects \\of tube constrictionsfor single loop PHP.\\ Many key phenomena are reproduced. \end{tabular}               \\ \hline

                                                                       \end{tabular}
                                                                       \end{table}
 \end{landscape}
 \newpage
 \begin{landscape}

\begin{table}[ht]
                                                                             \begin{tabular}{lllll}
                                                                             \hline
                                                                             Approach                                                              &References                                                                                                                         & \begin{tabular}[c]{@{}l@{}} Comparison\\ with experiment\end{tabular}                           &\begin{tabular}[c]{@{}l@{}} Validation\\ parameters\end{tabular}&Remarks  \\ \midrule

                                                                  &\citet{shafii1}                                                                                     & none &-                                                                                                                     & \begin{tabular}[c]{@{}l@{}} First model including properly evaporation-condensation.\end{tabular}             \\
     & & & &  \\
                                                                        & \citet{Holley05}                                                                                 &   none  &-                                                                                                                   & \begin{tabular}[c]{@{}l@{}} First model accounting for \\plug coalescence and bubble nucleation. \end{tabular}            \\
        	& & & &  \\
                                                                                    & \citet{Mameli12,MameliMST12}            &  good  &\begin{tabular}[c]{@{}l@{}}  multiple:\\liquid momentum\\ maximum tube temperature\\ equivalent thermal resistances \end{tabular}  &  \begin{tabular}[c]{@{}l@{}} Improvement of the model \cite{Holley05} by implementation \\of tube bends and two-phase heat transfer coefficient \\as function of heating \\regimes.\end{tabular}           \\
& & & &  \\
                                                         & \citet{Manzoni16}                                                                              &  good &     \begin{tabular}[c]{@{}l@{}} multiple:\\transient time\\ temperatures\end{tabular}                                                                                                                   & \begin{tabular}[c]{@{}l@{}}Lumped parameter model based on \cite{Mameli12}. The\\ assumption of saturated vapor is abandoned.\end{tabular}           \\
& & & &  \\
 \begin{tabular}[c]{@{}l@{}} (6) 1D hydrodynamics\\ with phase change\end{tabular}                                             & \citet{IJHMT10}                                                                                           & good & \begin{tabular}[c]{@{}l@{}}frequency and amplitude\\ of oscillations\end{tabular}                                                                                                                         &\begin{tabular}[c]{@{}l@{}}FEC model introduction \\for single branch PHP. \end{tabular}         \\
	& & & &  \\
                                                                        & Nikolayev \cite{JHT11,IarATE17,MST19,ISOPHP19} &  present work & \begin{tabular}[c]{@{}l@{}} multiple: \\ evaporator temperature\\ adiabatic tube wall temperatures\\ fluid pressures\\ liquid plug velocity\\liquid plug length\\temperature distribution\end{tabular}                                                                                                                         & \begin{tabular}[c]{@{}l@{}} FEC model implementation for multi-branch\\PHP: CASCO code.  \end{tabular}            \\
  	& & & &  \\
                                                                        & \citet{Daimaru17}                                                                               & average & single                                                                                                                         &  \begin{tabular}[c]{@{}l@{}}  FEC model for multi-branch PHP  \\ to study the start-up behavior with check valves. \end{tabular}      \\
  	& & & &  \\
                                                                        & \citet{Bae17}                                                                                       &  good & single                                                                                                                      &   \begin{tabular}[c]{@{}l@{}}   Experimental validation of the model \cite{Senjaya14}\\ that implements the spatial and \\ temporal variation of  liquid film thickness.   \end{tabular}          \\\bottomrule
  \end{tabular}
    \caption{PHP Theoretical Models overview } \label{ModelsTab}
  \end{table}
 \end{landscape}

\section{Experimental}

\subsection{Test cell}
\begin{figure} [ht]
\centering
	\includegraphics[scale=0.45]{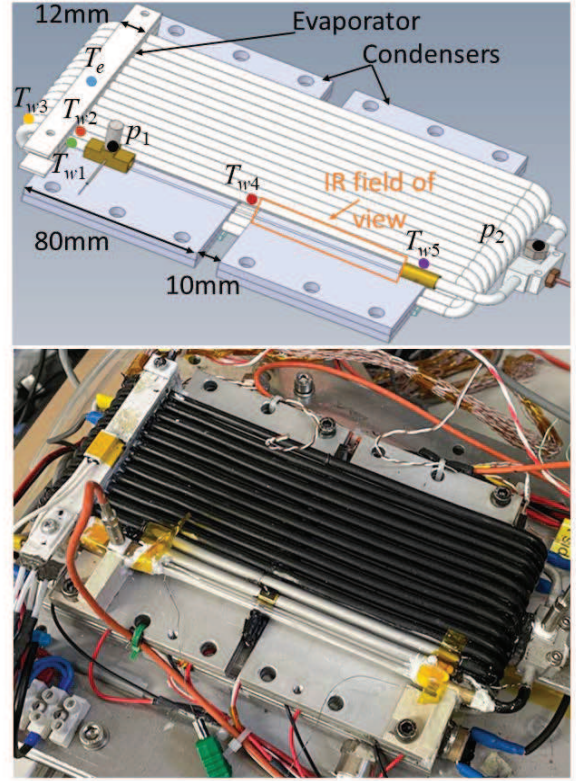}
	\caption{Test cell; Top: its drawing with thermocouple and pressure sensor locations and field of view of infrared camera. Bottom: photo.}
	\label{Testcell}
\end{figure}

The experimental device is one of the PHP prototypes designed to be implemented in the Heat Transfer Host 2 apparatus for the experiments on the International Space Station. The  device, shown in Figure~\ref{Testcell}, is a closed loop made of 6060 aluminum alloy tube (outer diameter $r_e=\SI{5}{mm}$, inner diameter $r=\SI{3}{mm}$) bent in 14 turns and arranged in a 3D structure. The evaporator zone consists of two aluminum spreaders ($100\times 12\times$\SI{10}{mm^3}) brazed on the tubes and heated with two ceramic heaters (Innovacera\textsuperscript{\textregistered}, electrical resistance $\SI{18}{\ohm}\pm 10$\%) powered by a programmable power supply (GW-Instek\textsuperscript{\textregistered}, PSH-6006A); the condenser is made of two aluminum heat spreaders ($80\times 120\times$\SI{10}{mm^3}) brazed on the tubes and kept at the desired temperature (with a maximum deviation of \SI{\pm 2}{K}) by using Peltier cells (by Adaptive Thermal Management\textsuperscript{\textregistered} ETH-127-14-11-S) and a control system (by Meerstetter  Engineering\textsuperscript{\textregistered}, TEC 1123) coupled with an external cold plate. The PHP is filled with 22\SI{\pm 0.2}{ml} of perfluorohexane (FC-72 by 3M\textsuperscript{\textregistered}) which provides the 50\% volumetric filling ratio. The tubes  in the evaporator zone and the heaters are thermally isolated with the polystyrene foam. Five T-type thermocouples  are located inside the evaporator block. Their temperatures are very close and therefore their average $T_e$ is taken into account. Next, there are five T-type thermocouples $T_{w1}\dots T_{w5}$ located on the external wall of an adiabatic tube sections (Fig.~\ref{Testcell}). There are eight T-type thermocouples to measure the Peltier and condenser spreader temperatures; their difference is also negligible so only their average value $T_c$ is considered. The fluid pressure is measured with two pressure transducers $p_1$ and $p_2$ shown in Fig.~\ref{Testcell} (Keller\textsuperscript{\textregistered} PAA-M5-HB, \SI{1}{bar} absolute, 0.2\% full scale output uncertainty). Finally, a 3-axis sensor (Dimension Engineering\textsuperscript{\textregistered}, DE-ACCM3d) is used for the measurement of the local accelerations. A data acquisition system (National Instruments\textsuperscript{\textregistered}, NI-cRIO-9074, NI-9264, NI-9214, 2xNI-9205, NI-9217, NI-9472) is connected to a laptop, and simultaneously acquires the thermocouples signal at \SI{50}{Hz}, the pressure transducers signal at \SI{200}{Hz} and $g$ at \SI{5}{Hz} via a LabView\textsuperscript{\textregistered} software.

A tube portion in the adiabatic section is replaced with a sapphire tube. A part of it (\SI{68}{mm}) is filmed with the Infrared camera (AIM\textsuperscript{\textregistered}, middle wave IR range, 3--\SI{5}{\mu m}) at the acquisition frequency $f=\SI{50}{Hz}$. The camera trigger is controlled by the LabView\textsuperscript{\textregistered} software and therefore synchronized with the rest of the above mentioned measurements.  From the image analysis, it is possible to assess the temperature \cite{Catarsi18,Perna20}, velocity and length of the liquid plugs appearing in the field of view. The uncertainties of the directly measured and derived quantities are summarized in Table~\ref{error}.
  \begin{table}[ht]
  \begin{tabular}{SS} \toprule
  {Parameter}  & 				{Uncertainty }        \\ \midrule
{$T_w, T_e, T_c$} & 				\SI{\pm 0.1}{K} \\
{$p_{1,2}$} & 				\SI{\pm 500}{Pa} \\
{$T_{sat}$} & 				\SI{\pm 0.5}{K} \\
$V$    &\SI{\pm 0.023}{m/s}\\
{$L_l$}  & \SI{\pm 0.5}{mm}                 \\
{$T_l$   systematic} & \SI{\pm 2}{K}                \\
{$T_l$   difference } & \SI{\pm 50}{mK}                \\ \bottomrule
  \end{tabular}
  \caption{Experimentally measured quantities and their uncertainties}\label{error}
  \end{table}
Note that the high uncertainty of the IR measurements of liquid plug temperature $T_l$ is systematic as shown in \citet{Catarsi18}; however, the temperature differences  are reproduced with a an accuracy of \SI{50}{mK}.

\subsection{Experimental procedure and available data}

\begin{figure} [ht]
\centering
	\includegraphics[scale=0.45]{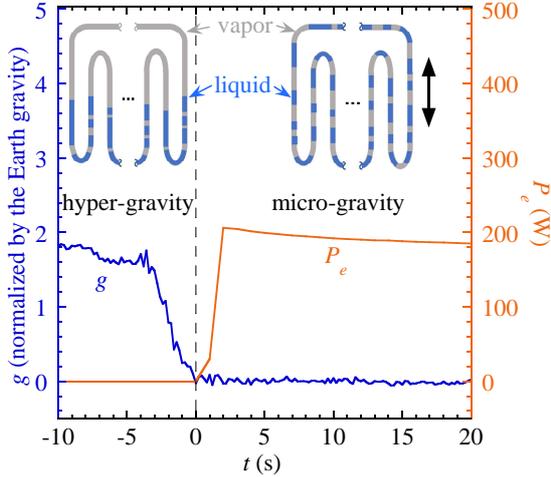}
	\caption{Evolution of the effective gravity acceleration $g$ and the evaporator power for parabola 16. An illustrative distribution of the bubbles and the plugs is shown above for both accelerated and microgravity intervals.}
\label{pattern}
\end{figure}

The experimental data used in the present work has been collected during the 67th ESA Parabolic Flight Campaign \cite{Mameli19}. During a parabolic flight, the airplane performs a series of parabolas (i.e.  maneuvers with a parabolic trajectory); each parabola is a sequence of a hyper-gravity period (20\SI{\pm 2}{s} where the effective gravity acceleration is about twice the Earth gravity), a microgravity period of parabolic trajectory (20\SI{\pm 2}{s} at no more than several percent of the Earth gravity) and again a hyper-gravity period (cf. Fig. \ref{pattern}) with a maximum interval between two parabolas of \SI{5}{min}.

For comparison of the transient behavior with numerical simulations, the initial state of the PHP should be well defined. Therefore, only the startup tests (i.e. those in which the device is heated up just after the microgravity occurrence and in which the device is initially in thermodynamic equilibrium with the environment) are chosen among all the available data. They are listed in Table~\ref{tableP}.

An example of the heating sequence adopted in the present work is shown in Figure~\ref{pattern}. During the first part of the parabolic maneuver ($t<0$) characterized by hyper-gravity (blue line), no heat input (orange line) is supplied to the evaporator. As soon as microgravity condition is reached ($g=0$ at $t=0$), the heating power is provided to the evaporator until the end of the microgravity period ($t\simeq20s$). Therefore, all the data reported here describes the evolution of various parameters during the start-up in microgravity conditions.

The experimental datasets (one per parabola) include the evolution of the tube wall temperatures ($T_{w1}\dots T_{w5}$), the evaporator temperature ($T_e$)  and the pressures in terms of saturation temperature $T_{sat}$ during the \SI{20}{s} microgravity period.  During some parabolas, liquid plugs are visible in the field of view (IR column of Table~\ref{tableP}) and the data on the liquid plug velocity, length and temperature distribution obtained from the analysis of the IR images are added to the corresponding dataset. A simultaneous comparison on all these parameters which characterize the PHP dynamics is thus performed.

\begin{table}[ht]
\begin{tabular}{ccccc} \toprule
Parabola \#   &$P_e$ [\si{W}]&$q_e$ [\si{kW/m^2}]& { IR  }&  {Deviation [\%]} \\ \midrule
2             & 35               &13                       & {-}    & 4                     \\                              
16            & 205              &77                       & {-}    & 7               \\                                  
19            & 200      &76                       & {yes } & 5                  \\                                
22            & 135              &51                       & {-}    & 5                   \\                                  
25            & 135              &51                       & {-}    & 4                  \\                                   
27            & 70               &26                       & {-}    & 4                    \\                               
30            & 70               &26                       & {yes}  & 3                  \\ \bottomrule                          
\end{tabular}
\caption{Experimental datasets and maximum temperature deviation between experiment and simulation (see sec. \ref{TPsec}).}\label{tableP}
\end{table}

\section{Simulation setup}

Since the CASCO equations are dispersed over several publications \cite{JHT11,IarATE17,IHPC18VN,MST19,ATE21}, the full model is presented in \ref{AppFEC}.

\begin{figure}
  \centering
  \includegraphics[width=0.4\textwidth]{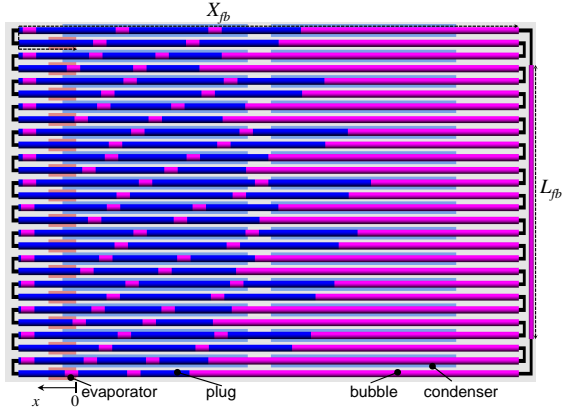}
  \caption{2D representation of the experimental prototype (Fig. \ref{Testcell}) by CASCO software and the initial liquid-vapor distribution inside the PHP. The round turns are not represented for simplicity; black lines
are simply connectors linking the equivalent points of neighboring
branches. Light blue and light red rectangles identify the condenser and evaporator sections, respectively. Thin liquid films (in violet) cover entirely the internal tube walls
inside the vapor bubbles. The liquid plugs are blue.}\label{Init}
\end{figure}

To provide a truthful simulation, the CASCO input data (Table~\ref{CascoInput}) should faithfully reproduce the actual PHP in terms of geometry, topology, material properties and boundary conditions. In particular, CASCO accounts for two distinct heat sinks spaced by a small adiabatic section (Fig.~\ref{Testcell}). There are two more adiabatic sections: one between a condenser and the evaporator and another between the evaporator and the other condenser.
The 3D PHP structure projected to a 2D plane is shown in Figure~\ref{Init}. The branches containing an evaporator section alternate with the branches containing two condenser sections.

The fluid properties are evaluated with NIST REFPROP\textsuperscript{\textregistered} 9. The liquid viscosity and $p_{sat}$ are considered to be functions of temperature while all the other quantities are taken as constants (at \SI{50}{^\circ C}) because their temperature variation is much weaker.

The adiabatic boundary conditions are defined at the external surface of the evaporator block (called spreader hereafter). The power $P_e$ injected into it follows the experimentally measured time dependence for each parabola. The condenser temperature $T_c$ is assumed to be constant.

The spreader thermal mass $C_s\simeq \SI{67}{J/K}$ and the contact conductance $U_s\simeq\SI{1600}{W.m^{-2}.K^{-1}}$ between the heater and the tube where obtained by fitting of the calculation results to the experimental data for the case of the empty PHP.

The initial liquid phase distribution inside the tube is unknown; one knows only what happens in one transparent branch. However, it is possible to make some assumptions based on the knowledge of the PHP state before the start-up. During the hypergravity part of the parabolic maneuver preceding the PHP startup ($t<0$), the effective gravity acceleration is directed towards the evaporator. Therefore, based on previous observations in similar devices \cite{Ayel19}, it is assumed that, initially, all the liquid gathers in the evaporator side along with some randomly distributed small bubbles in the plugs. Arbitrarily (the global PHP evolution does not depend much on these parameters), five small bubbles are distributed homogeneously in each plug occupying the total volume fraction of 0.06. The initial plug distribution (Fig.~\ref{Init}) is chosen so the liquid is not initially visible in the IR field of view, which is experimentally observed for all the parabolas.

The liquid film thickness $\delta_f$ (Table~\ref{CascoInput}) is calculated with the \citet{Aussillous} formula. The parameters of FC-72 are taken at \SI{26}{^\circ C} and the meniscus velocity is \SI{0.15}{m/s}. These two values is a result of an iteration procedure and are obtained as follows. First, a reasonable $\delta_f$ value is chosen as an input and CASCO is run. The average temperature and plug velocity that occurred during evolution are calculated. Their values are used to compute $\delta_f$ with the \citeauthor{Aussillous} formula. The procedure is repeated until the resulting $\delta_f$ coincides with its input value within a reasonable accuracy. One needs usually several (3-4) iterations to achieve convergence.

The nucleation barrier $\Delta T_{nucl}$ (i.e. the minimum wall superheating required for the bubble nucleation) is used as a tuning parameter. A chosen value  $\Delta T_{nucl}=\SI{4.3}{K}$ (Table~\ref{CascoInput}) agrees with \citet{Wang12} where $\Delta T_{nucl}$ is shown to be of the order of \SI{10}{K} for a mass flux around \SI{35}{kg.m^{-2}.s^{-1}} and a heat flux of the order of \SI{10}{kW.m^{-2}} (i.e. approximately mass and heat flux of the experimental test cell).

The results of the simulations performed with such a set-up are compared hereafter to the experimental data.

 \begin{table}[ht]
 \begin{tabular*}{\columnwidth}{@{\extracolsep{\fill}}l@{\extracolsep{\fill}}l@{\extracolsep{\fill}}}
 \toprule
Parameter and its notation& value\\
\midrule
Number of turns, $N_{turn}$                                    &           14\\
Number of hot sources, $N_e$                                    &           1\\
Number of cold sources, $N_c$                                    &           2\\
Length of the hot zone, $L_e^1$                    &           \SI{12}{mm}\\
Lengths of the cold zones, $L_c^k$                         &           \SI{80}{mm}, \SI{80}{mm}\\
Lengths of the adiabatic zones, $L_a^k$                        &           \SI{32}{mm}, \SI{10}{mm}, \SI{22}{mm}\\
Length of the feed-back section, $L_{fb}$      &           \SI{84}{mm}\\
Feed-back offset, $X_{fb}$           &           \SI{241}{mm}\\
Tube inner radius, $r$                                                         &           \SI{1.5}{mm}\\
Tube outer radius, $r_e$                                           &           \SI{2.5}{mm}\\
Tube bend radius        &           \SI{8}{mm}\\
Filling ratio                                                         &           0.5\\
Initial temperature                           &           \SI{21}{^\circ C}\\
Condenser temperature, $T_c$                                  &           \SI{20}{^\circ C}\\
Time step                                                              &           \SI{0.1}{ms}\\
Wall element length                                                &           \SI{2}{mm}\\
Liquid element length                                     &           \SI{1}{mm}\\
Nucleated bubble length, $L_{nucl}$                                  &           \SI{100}{\mu m}\\
Nucleation distance, $L_{nucl,min}$         &           \SI{5}{mm}\\
Nucleation barrier, $\Delta T_{nucl}$                          &           \SI{4.3}{K}\\
Bubble deletion threshold, $L_v^{thr}$                                  &           \SI{10}{\mu m}\\
Plug deletion threshold, $L_l^{thr}$                                   &           \SI{2}{mm}\\
Liquid film thickness, $\delta_f$                                                             &  \SI{72.3}{\mu m}        \\
\bottomrule
\end{tabular*}
\caption{CASCO input parameters (cf. \ref{AppFEC}). The material parameters for the fluid and the tube material are taken for \SI{50}{^\circ C}.}\label{CascoInput}
\end{table}

\section{Results: comparison of experiment and simulation}
\subsection{Tube and evaporator temperatures}\label{TPsec}

\begin{figure}[ht!]
\begin{subfigure}{.45\textwidth}
\centering	\includegraphics[scale=0.3]{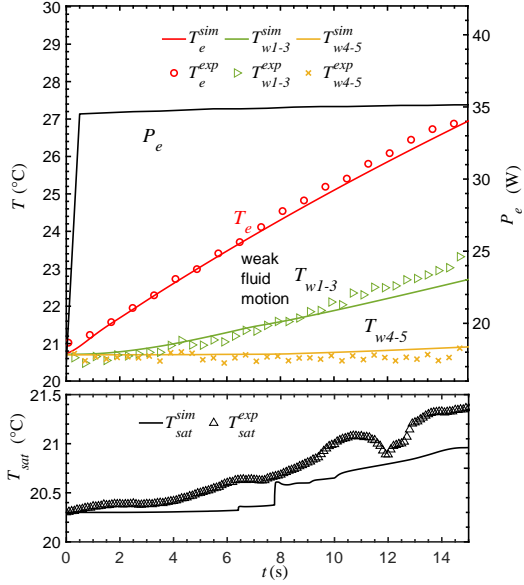}
	\caption{Parabola 2 data.}
	\label{temps2}
\end{subfigure}
\begin{subfigure}{.45\textwidth}
\centering	\includegraphics[scale=0.3]{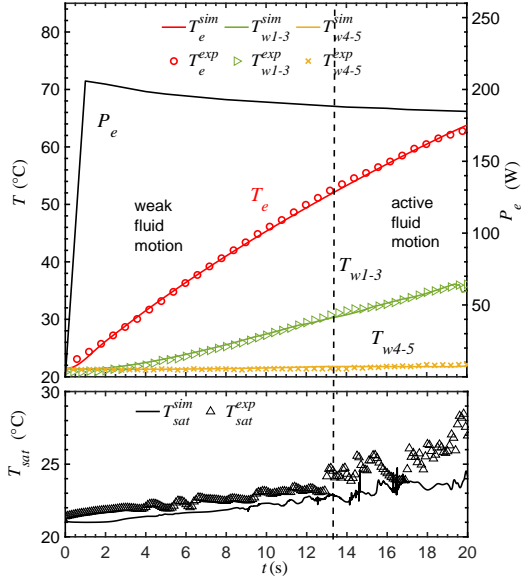}
	\caption{Parabola 16 data.}\label{temps3}
\end{subfigure}
  \caption{Evolution of evaporator, tube wall and saturation temperatures for parabolas 2 and 16. Experiment: characters; simulation: lines.}\label{T1}
\end{figure}

\begin{figure}[ht!]
\begin{subfigure}{.45\textwidth}
\centering	\includegraphics[scale=0.3]{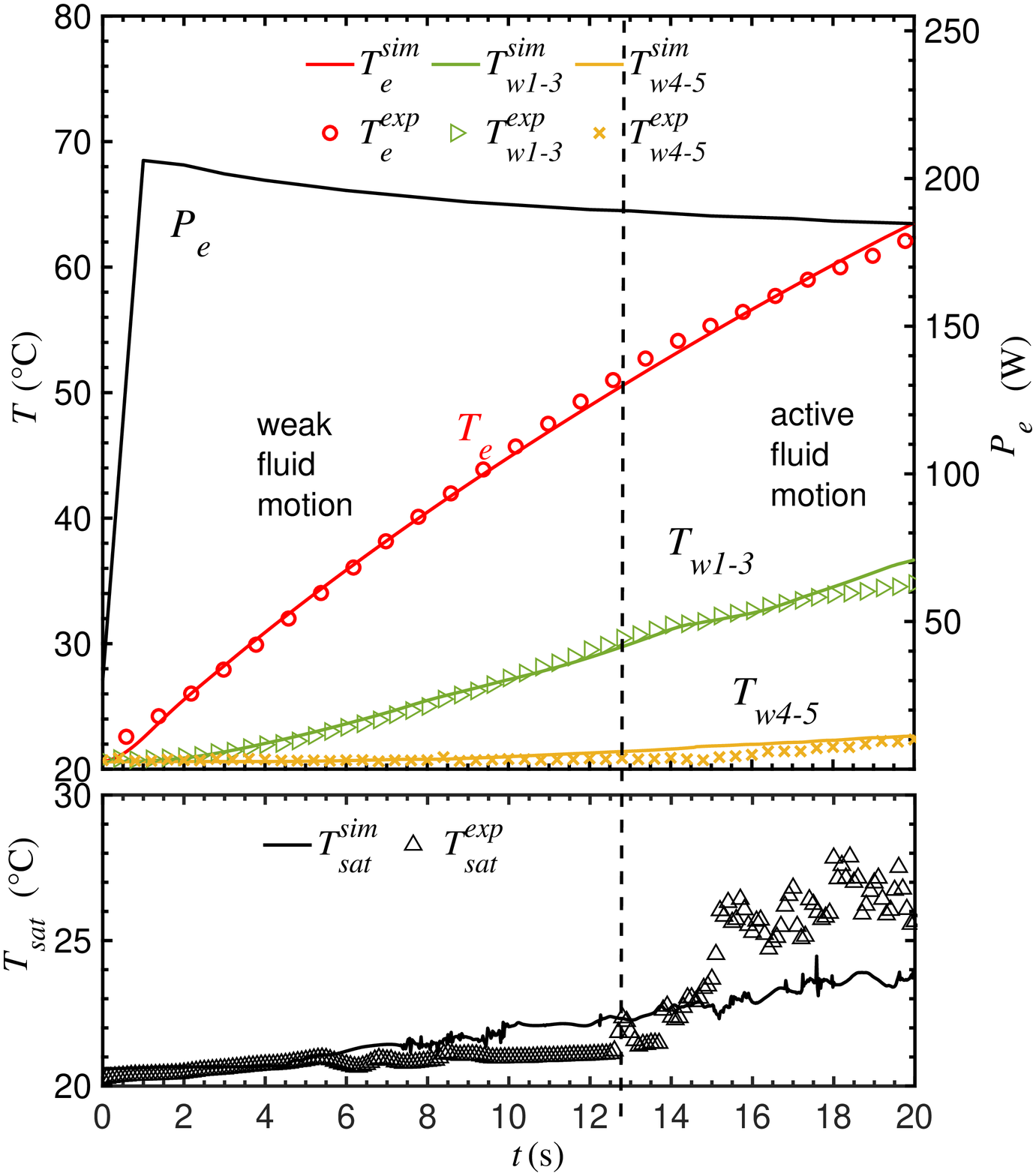}
	\caption{Parabola 19 data.}\label{temps}
\end{subfigure}
\begin{subfigure}{.45\textwidth}
\centering	\includegraphics[scale=0.3]{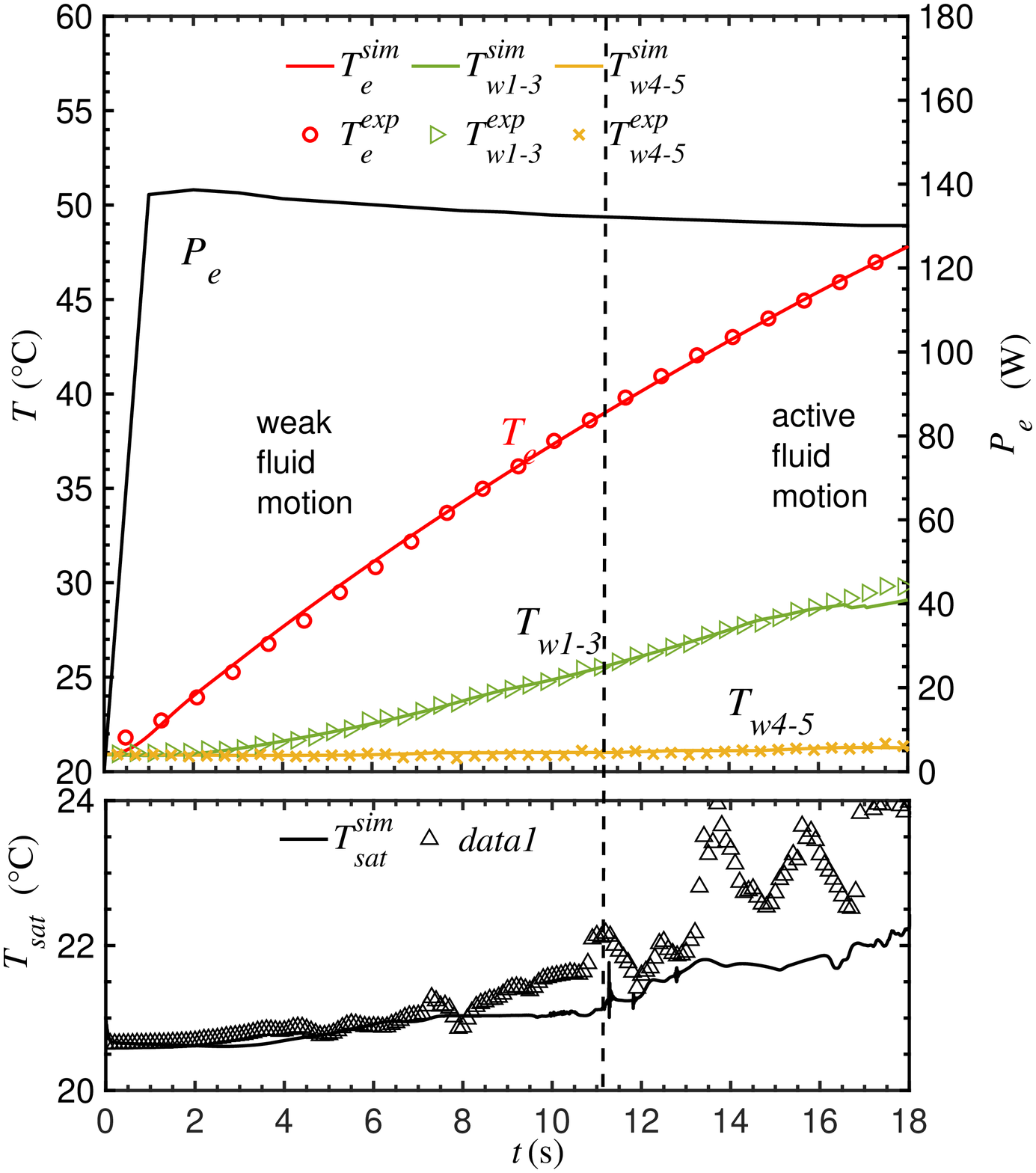}
	\caption{Parabola 22 data.}
	\label{temps22}
\end{subfigure}
  \caption{Evolution of evaporator, tube wall and saturation temperatures for parabolas 19 and 22. Experiment: characters; simulation: lines.}
\end{figure}

\begin{figure}[ht!]
\begin{subfigure}{.45\textwidth}
\centering		\includegraphics[scale=0.3]{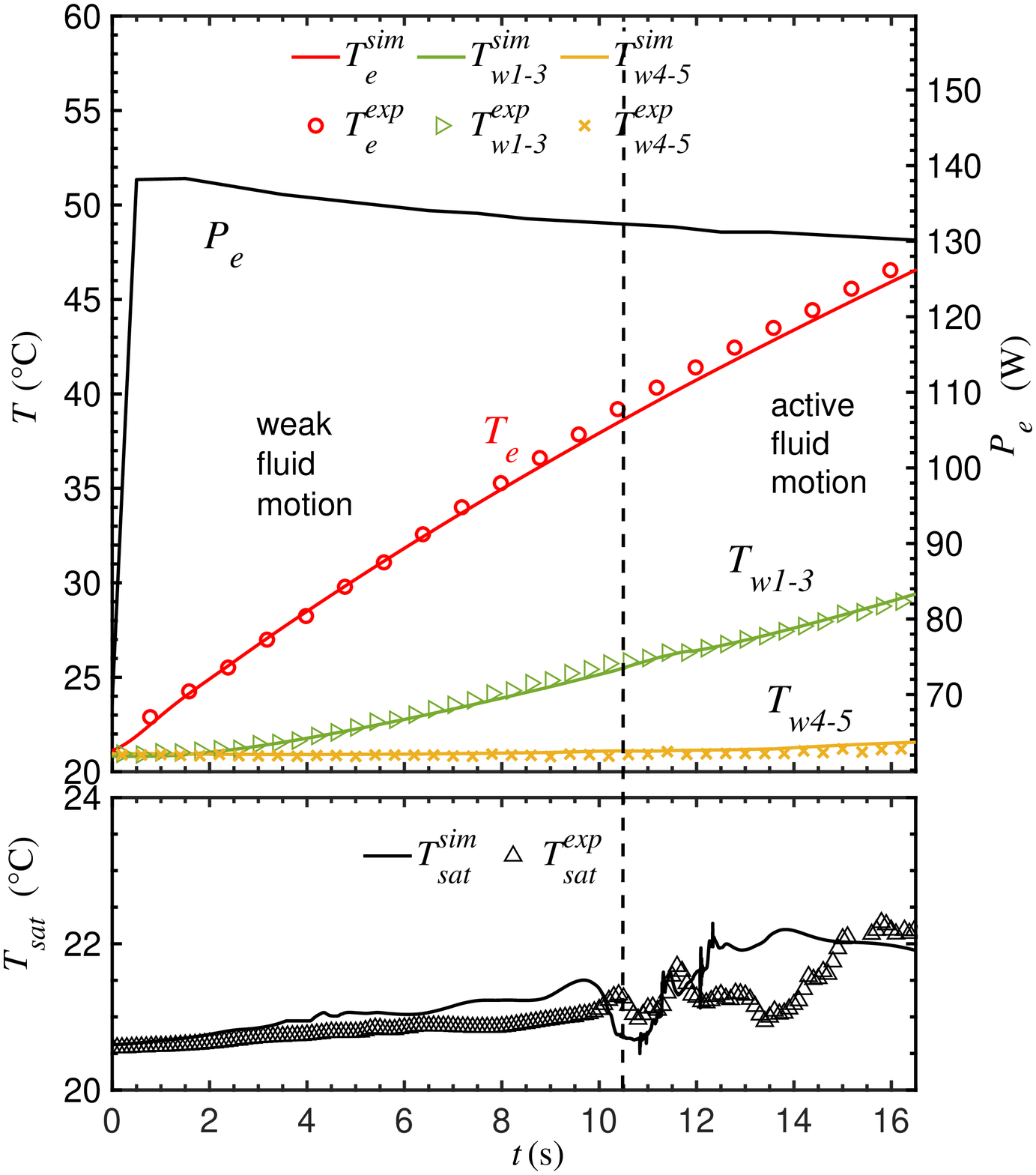}
	\caption{Parabola 25 data.}\label{temps25}
\end{subfigure}
\begin{subfigure}{.45\textwidth}
\centering	\includegraphics[scale=0.3]{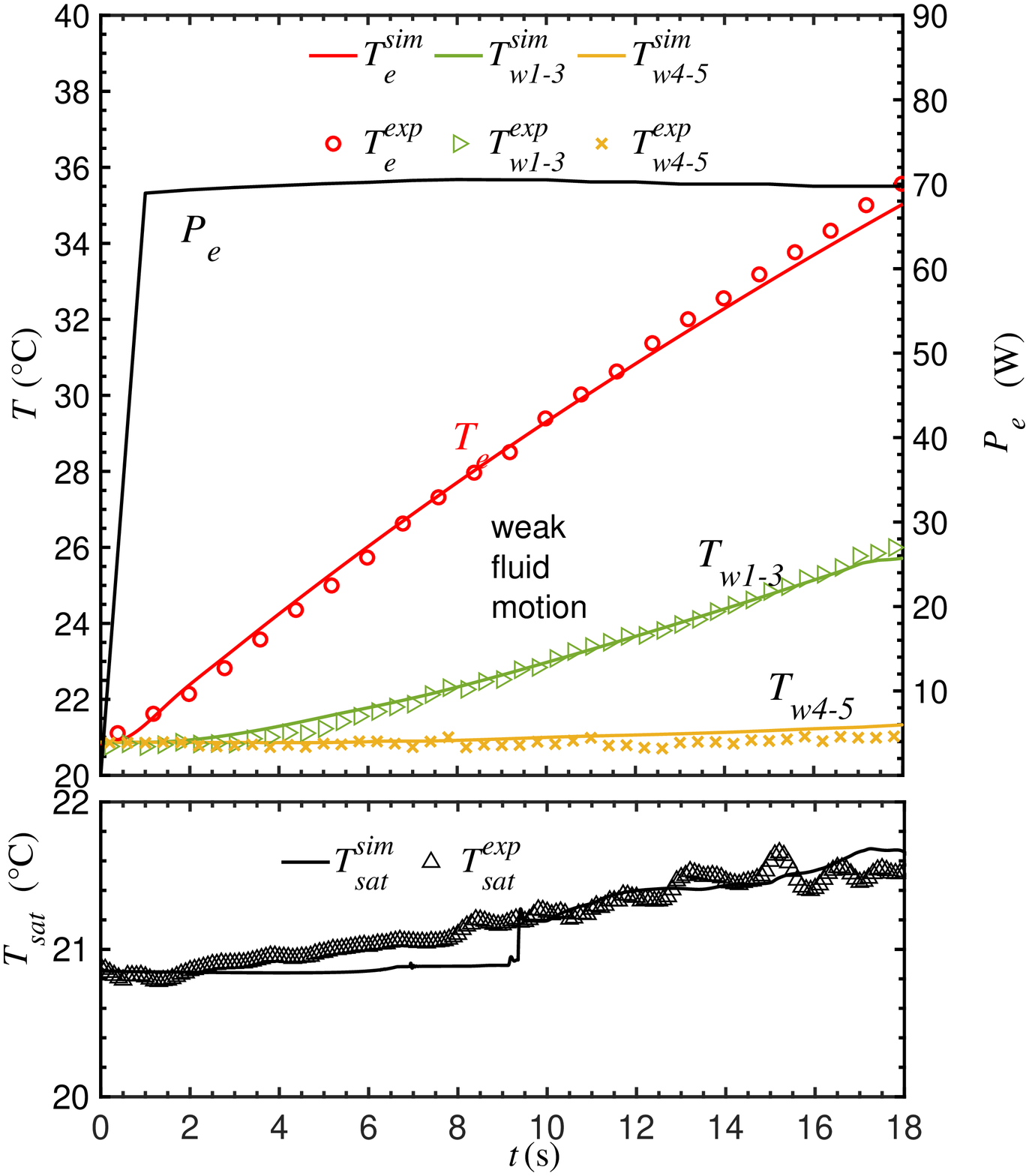}
	\caption{Parabola 27 data.}\label{temps27}
\end{subfigure}
  \caption{Evolution of evaporator, tube wall and saturation temperatures for parabolas 25 and 27. Experiment: characters; simulation: lines.}
\end{figure}

\begin{figure}[ht!]
\centering	\includegraphics[scale=0.3]{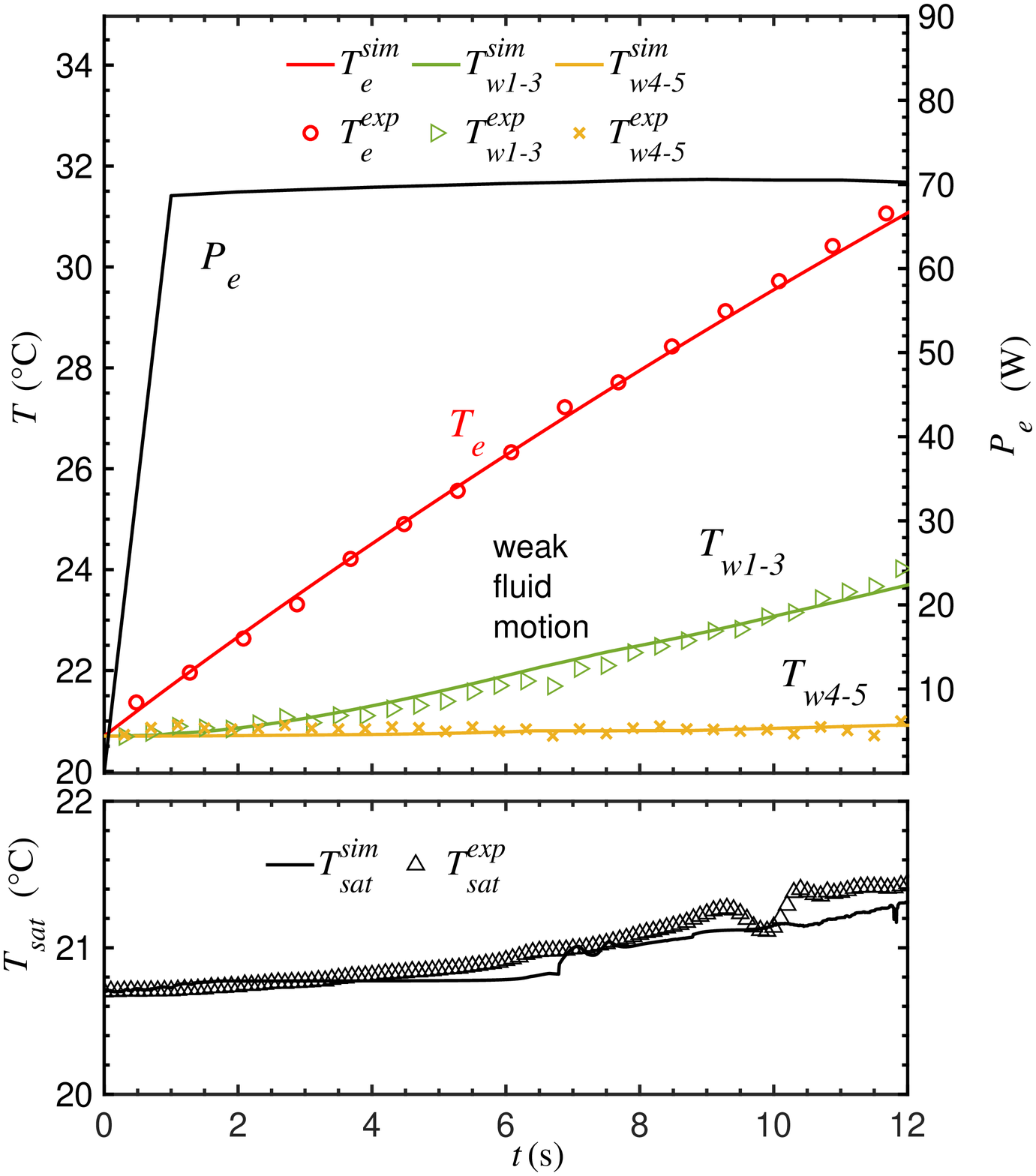}
	\caption{Evolution of evaporator, tube wall and saturation temperatures for parabola 30. Experiment: characters; simulation: lines.}\label{temps30}
\end{figure}

The temporal variations of tube and evaporator temperatures are compared with simulations in Figs.~\ref{T1}--\ref{temps30} for different parabolas. The positions along the tube where the simulated temperatures $T_{w1}\dots T_{w5}$ are recorded correspond to the locations of the thermocouples 1\dots 5 shown in Figure~\ref{Testcell}. Similarly, the pressures $p_1,\,p_2$ are recorded in simulation in the precise experimental locations of the transducers 1 and 2. The thermocouples 1\dots 5 can be divided into two groups. The thermocouples 1\dots 3 are close to the evaporator, while the thermocouples 4, 5 are close to the condenser, see Figure \ref{Testcell}. Accordingly, for the sake of clarity of the Figures~\ref{T1}--\ref{temps30}, only the averaged values $T_{w1-3}=(T_{w1}+T_{w2}+T_{w3})/3$ and $T_{w4-5}=(T_{w4}+T_{w5})/2$ are presented. The pressure variation is presented for each experiment in terms of the saturation temperature $[T_{sat}(p_1)+T_{sat}(p_2)]/2$ that can be compared with the wall temperatures. In the experiment, as soon as the microgravity occurs, the power is injected into the evaporator spreader, whose temperature $T_e$ starts to increase. After a few seconds, the tube wall temperature increases too as a result of heat diffusion from the evaporator zone. Higher wall temperatures occur close to the evaporator ($T_{w1-3}$ plotted in green) and lower temperatures, in the locations close to the condenser ($T_{w4-5}$ plotted in yellow). Both simulation and experiments show that the wall temperature in the evaporator  $T_e>T_{sat}$ causes the film evaporation, and expansion of some of already existing bubbles; others collapse because they are compressed by the fast plug motion. When the local wall superheating $\Delta T =T_w-T_{sat}$ reaches $\Delta T_{nucl}$, new bubbles are nucleated and oscillations start. The graphs show that this occurs quite soon both in the evaporator area and close to it. Progressively, the bubble nucleation and growth lead to the active plug motion detected in the transparent section (see sec. \ref{Vsec} below).


Fig.~\ref{temps25} refers to \SI{\sim 135}{W} power input. By looking at pressure evolution, it seems that the starting time of fast fluid motion is not strongly influenced by the power level. In fact, the first pressure perturbation appears around \SI{11}{s} which is close to the cases of high power \SI{\sim 200}{W} (Figs.~\ref{temps3}, \ref{temps}). On the other hand, as a result of lower temperatures (and then less frequent appearance of dry spots in the evaporator sections), the pressures strongly oscillate and the $T_{sat}$ rise appears less pronounced (only \SI{\approx 1}{K} between \SI{10}{s} and \SI{15}{s}) as compared to higher $P_e$ cases (\SI{\approx 4}{K} between \SI{14}{s} and \SI{19}{s} in the cases of Figure~\ref{temps3} and Figure~\ref{temps}).

For the sake of completeness, the results obtained at low heat loads are also reported. As shown in  Figs.~\ref{temps2}, \ref{temps27} and  \ref{temps30},  the power input is not large enough to produce any notable effect in terms of pressure and, consequently, in terms of saturation temperature. Similarly, tube wall temperature rise is predominately the result of heat diffusion through aluminum tubes and it is not strongly altered by interaction with working fluid. In fact, fluid motion is characterized by a low velocity as result of slow expansion of bubbles in the evaporator zone as confirmed by fluid visualization in the case of Figure~\ref{temps30} below.

In order to quantify the prediction ability of CASCO, the simulation accuracy is evaluated using a formulation similar to that adopted by \citet{Wang20}.

 \begin{equation}
 \mathrm{deviation}=\max \bigg \{ \frac{|T^{exp}(t)-T^{sim}(t)|}{0.5 (T^{exp}(t)+T^{sim}(t))}\bigg \} \times 100
 \label{dev}
 \end{equation}
where $T^{exp}(t)$ and $T^{sim}(t) $ are respectively experimental and simulation temperatures as function of time; results are shown in Table \ref{tableP} for each parabola.
The maximum deviation occurs for the \SI{\sim 205}{W} power input case plotted in Figure~\ref{temps3} between $T_{w1}^{exp}$ and $T_{w1}^{sim}$ at \SI{\sim 20}{s}. It is however evident that the simulation is able to capture the start-up temperature trend in all the experimentally investigated conditions. The case of parabola 2 (Fig.~\ref{temps2}) appears however as the least accurate case in terms of absolute deviation value. The origin of this deviation is evident from the comparison of the initial temperature values. There is almost a \SI{1}{K} difference between the wall and saturation temperatures, which means that the fluid is not initially isothermal (contrary to what was assumed in the simulation). Because of the small $P_e$, this difference is large with respect to the overall $T_{sat}$ rise.

Pressure trends (reported in terms of saturation temperature) are qualitatively reproduced. Some minor discrepancy occurs due to the intrinsically stochastic nature of the phenomena influencing pressures evolution (i.e. initial liquid distribution,location of bubbles nucleation spots, presence of liquid film etc.; all of these make almost impossible to reproduce exactly. Therefore, in order to obtain an indication of model ability of capturing the actual fluid behavior, a local-level analysis is needed; it is done by comparing liquid plug characteristics.

\subsection{Liquid plug velocity and length}\label{Vsec}
\begin{figure} [!ht]
\begin{subfigure}{.45\textwidth}
\centering\includegraphics[scale=0.3]{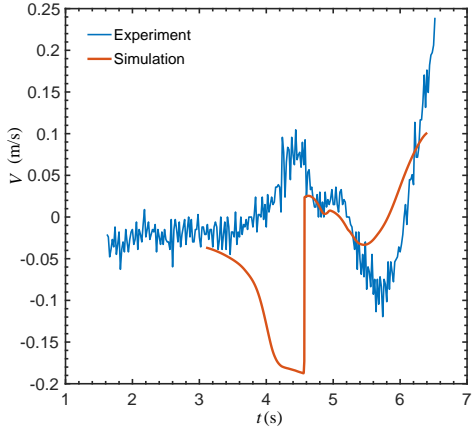}
\caption{Parabola 19 data.}\label{vel}
\end{subfigure}
\begin{subfigure}{.45\textwidth}
\centering\includegraphics[scale=0.3]{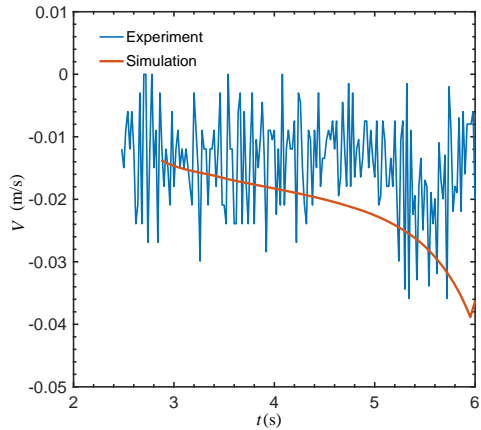}
	\caption{Parabola 30 data.}\label{vel30}
\end{subfigure}
\caption{Temporal evolution of liquid plug velocity during two parabolas. Experimental velocity is in blue while the simulated velocity is in orange.}\label{VEL}
\end{figure}

First, the liquid plug velocity is analyzed; this is a quantity rarely reported in the PHP literature. It is obtained from the analysis of IR images captured through sapphire tube section partially transparent for the IR radiation. The image processing for plug recognition and tracking is thoroughly described in \cite{Perna20}. Here, only some basics are reminded. A  Matlab\textsuperscript{\textregistered} script is used to systematically analyze the images acquired by the infrared camera. The script is able to detect the liquid plugs menisci locations. Once detected, each liquid plug is recognized in the consecutive movie frames as a plug of the close length. In this way, it is possible to track each plug and evaluate its velocity. It follows that the velocity of  a plug $i$ visible in the $k$-th time step is equal to $V_{i,k}=(x_{i,k}-x_{i,k-1})f$, where $x_{i,k}$ is the coordinate of its center of mass on the $k$-th time step,  $x_{i,k-1}$ is the coordinate of its center of mass on the $k-1$-th time step and $f$ is the acquisition frequency. Since the transparent section is adiabatic, no or little phase change takes place; therefore, liquid plugs move as a \emph{train} (i.e. with almost the same velocity). Consequently, the \textit{train} velocity is sufficient to describe their motion and it is calculated as
 \begin{equation}
 V^{exp}_k=\frac{1}{n_k}\sum V_{i,k}.
 \end{equation}
Such averaging is used to reduce the uncertainty related to the definition of the menisci. Figures~\ref{VEL} show the experimental and simulation liquid plug velocities for parabolas 19 and 30.
In the case of Fig.~\ref{vel} (\SI{\sim 200}{W} power input), simulation is able to reproduce quite accurately the experimental trend; peaks location is correctly predicted as well as the amplitude. The case of Figure~\ref{vel30} refers to a much lower power input (\SI{\sim 70}{W}) which produces a less intensive motion. In this case, during the first \SI{5.5}{s}, the simulation correctly predicts the motion characterized by a low negative velocity. After \SI{5.5}{s}, a difference appears. In simulation, the initial plug motion is driven by the expansion of bubbles within the plugs that cover initially the evaporator. Its velocity $V$ is directed from evaporator toward the condenser, i.e $V<0$ in the branch visible by the camera (the lowest in Fig.~\ref{Init}, where the positive $x$ direction is shown), which agrees with the flow direction observed in parabola 30 (Fig.~\ref{vel30}). A different initial $V$ sign is observed experimentally in parabola 19 (Fig.~\ref{vel}), which suggests an initial temperature difference between the fluid in the evaporator and the feed-back section that cannot be foreseen in simulation because it is unknown. For both parabolas, CASCO is however able to predict the experimental behavior in terms of temporal location and amplitude of the velocity peaks.

\begin{figure} [!ht]
\begin{subfigure}{.45\textwidth}
\centering\includegraphics[scale=0.3]{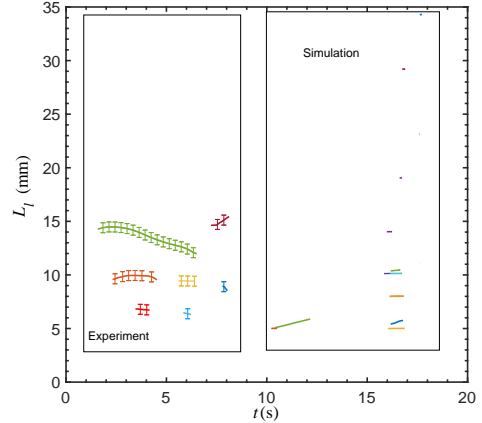}
\caption{Parabola 19 data.}\label{length}
\end{subfigure}
\begin{subfigure}{.45\textwidth}
\centering\includegraphics[scale=0.3]{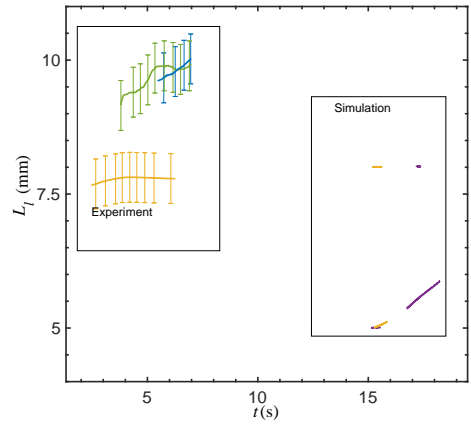}\caption{Parabola 30 data.}\label{length30}
\end{subfigure}
\caption{Evolution of the length $L_l$ of visible liquid plugs during two parabolas. Experiment: triangles; simulation: circles.}\label{LEN}
\end{figure}

Figs.~\ref{LEN} show a comparison of liquid plug lengths. They are calculated in a similar way in experiment and in simulation. Each color corresponds to a specific liquid plug appearing in the view field during the start-up. Despite the fact that liquid plugs appear later in simulation with respect to the experiment, there is a good agreement of their lengths. A time shift between the first recorded simulation length and first recorded simulation velocity is due to the fact that the length can be evaluated when both menisci are in the field of view while the velocity can be evaluated when at least one meniscus is in the field of view.
While the liquid plug stays visible in the transparent section, its length can decrease or increase depending on its position and velocity. If the plug velocity is large, residence time in the transparent section is small, so its length cannot change considerably. Therefore, the short lines in Figs.~\ref{LEN} are horizontal. If the plug velocity is small so the plug remains for a longer time inside the transparent section (long lines in Figs.~\ref{LEN}), its length depends on its position. If, most of the residence time, a plug stays closer to the evaporator side of the section, its length decreases because of evaporation. Inversely, if a plug stays closer to the condenser side, its length increases. If it displaces between the ends, its length derivative can vary.

\subsection{Liquid plugs temperature distribution}

\begin{figure} [!ht]
\begin{subfigure}{.45\textwidth}
\centering\includegraphics[scale=0.4]{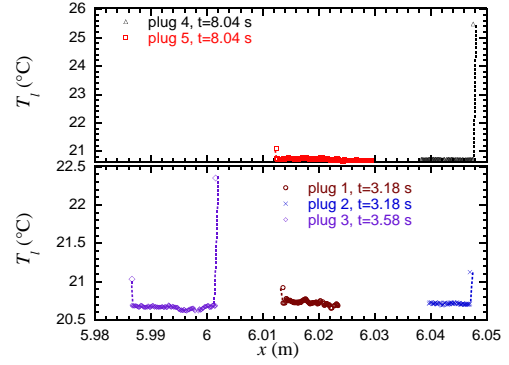}
	\caption{Experimental distributions. The lines are put as an eye guide. The plug velocities are 0.018, 0.018, 0.018, 0.065 and \SI{0.065}{m/s} for the plugs 1--5, respectively. $T_{sat}(p_2)$ and $T_{sat}(p_1)$ taken at the respective time moments are added as lower and upper bounds, respectively, for the interfacial values at the plug menisci.}\label{distr}
\end{subfigure}
\begin{subfigure}{.45\textwidth}
\centering\includegraphics[scale=0.4]{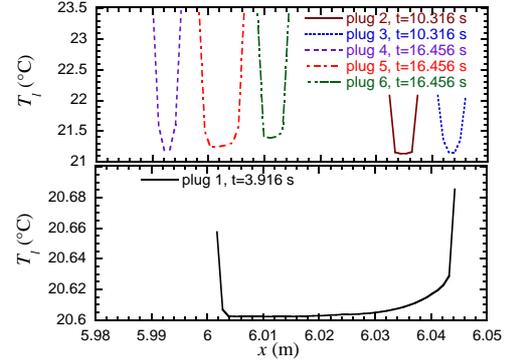}
\caption{Simulation. The plug velocities are -0.101, -0.113, , -0.121, 0.034, 0.022 and \SI{0.015}{m/s} for the plugs 1--6, respectively.}\label{distrSim}
\end{subfigure}	
\caption{Compared experimental and simulated temperature distribution inside liquid plugs during parabola 19.}
\label{Tliq}
\end{figure}

Finally, a comparison of liquid temperature distributions is performed. Since sapphire is transparent to the middle-wave IR radiation, it is possible to capture small temperature gradients within the liquid phase (vapor IR emission is too weak to be detected) and to catch temperature distributions of relatively fast thermo-fluid dynamic events. It is a prominent ability of the present experiment.  By using the method described by \citet{Catarsi18}, the IR camera is calibrated by varying the fluid and the ambient temperatures in a thermal chamber; in this way, the emitted IR radiation is linked to the back-screen temperature and the fluid temperature. Moreover, a lumped parameter radiation model is developed to quantify the effect of the involved parameters (ambient temperature, back screen temperature and its emissivity, tube transmissivity, fluid transmissivity etc.), in case the experimental conditions differ from the calibration. As result, the temperature distribution of liquid plugs is measured with a maximum error of \SI{\pm 2}{K} (Table \ref{error}). In Figures~\ref{Tliq}, the experimental and simulated distributions of liquid temperature $T_l(x)$ are presented at several time moments. At the plug menisci, the liquid temperature is equal to the saturation temperatures corresponding to the pressures in the preceding and the next vapor bubbles, respectively. While they cannot be measured experimentally with certainty, they can be reasonably estimated by the values $p_2$ and $p_1$ (in the order of $x$ increase) given by the respective pressure transducers that situate on the both sides of the visible section (Fig.~\ref{Testcell}). The characters corresponding to $T_{sat}(p_2)$ and $T_{sat}(p_1)$ at the respective time moments are thus added to Figure~\ref{distr} at the positions of plug menisci. They are linked to other characters of the same plug with dashed lines. It should be noted the transducers 1 and 2 can be outside the vapor bubbles neighboring the observed liquid plug. Note that $T_{sat}(p_2)$ and $T_{sat}(p_1)$ are only the bounds (and not the exact values) for the interfacial temperatures so their difference can be larger than in the reality. When there is a unique plug in the field of view, the estimations can be given for the temperatures of both menisci (plug 3 in Fig.~\ref{distr}). When there are simultaneously two plugs, the estimations can be given only for leftmost and rightmost visible menisci (plugs 1,2 and plugs 4,5 in Fig.~\ref{distr}). From the $p_2$ and $p_1$ comparison it is clear that the velocity of all the plugs should be negative in the respective time moments, which conforms to the velocity measurements shown in Fig.~\ref{vel}.

The comparison of the experimental (Fig.~\ref{distr}) and simulation (Fig.~\ref{distrSim}) results show very similar $T_l(x)$ distributions with sharp thermal boundary layers near the menisci. They appear because the change of bubble pressure can be much faster than that of the liquid temperature. Indeed, while the latter is controlled by the thermal diffusion, the bubble pressure is controlled by the superposition of change in the dynamics of a large number of plugs inside the PHP so sharp variations are statistically frequent.  Accounting for a relatively large systematic error on the temperature (Table~\ref{error}), the agreement between experimental and simulation liquid temperature distribution is very good. Due to the low power input, the parabola 30 liquid temperature measurement is highly affected by the noise (cf. Fig.~\ref{vel30}) and cannot thus be used for $T_l(x)$ measurements.

\section{Steady state simulation}
\begin{figure} [ht]
	\centering\includegraphics[scale=0.3]{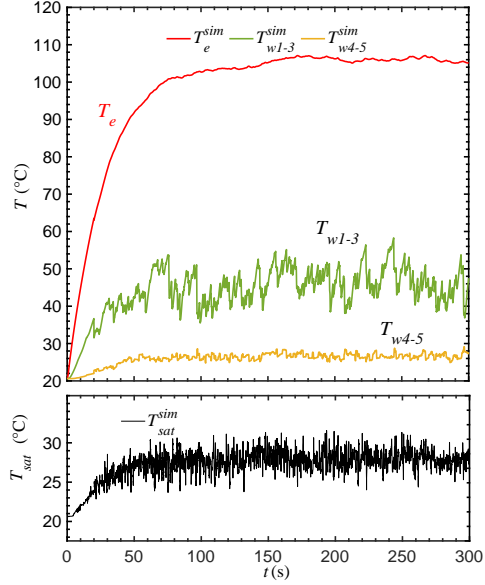}
	\caption{Temperature and pressure evolution of evaporator thermocouples and tube thermocouples in  run.}
	\label{longtemps}
\end{figure}
In Fig.~\ref{longtemps} we present the CASCO predictions for the long-time PHP functioning of the parabolic flight prototype. First \SI{20}{s} of the functioning coincide with Fig.~\ref{temps}. After this interval, a constant $P_e=\SI{185}{W}$ is assumed. One can see that the steady (or, rather, pseudo-steady) state is established after only \SI{1}{min} of functioning, which demonstrates high PHP heat transfer rate. A very efficient continuous oscillation regime with no stopovers is observed. One can clearly see the effect of evaporator spreader that smoothes the thermal fluctuations of the tube wall temperatures $T_{w1-3}$ measured in the evaporator vicinity (however already smoothed by averaging). Note a high difference between the spreader and wall temperature appeared because of low contact conductance $U_s$ (probably due to the imperfection of welding of the spreader to the tubes), which was deduced from the experimental thermal analysis of the empty PHP.
These results will be validated against the experimental data provided by the ISS experiment scheduled in the coming years. If the device will behave as predicted by the numerical long term simulation, the assessment of a novel large diameter PHP for space applications will be complete.

\section{Conclusions}

Numerical models are indispensable tools for the optimization of PHP design. In this work, the CASCO code prediction ability is validated against the experimental data collected during the micro-gravity tests of an innovative large diameter PHP designed to be implemented on the HTH apparatus onboard the International Space Station.
The CASCO software is used to accurately reproduce the actual device (in terms of geometry and topology) and test environment (imposing same initial and boundary conditions of the experiments). Simulations are run for the different power levels; their results are compared with the experimental data showing that the model is able to predict the device behavior not only globally but also at a local level.
\begin{itemize}
\item Temperature temporal evolution of evaporator and tube walls is closely captured with a maximum deviation of 7\%;
\item Pressures trends are qualitatively reproduced; simulation is able to reproduce sudden pressure variation observed experimentally indicating that fluid dynamics are well modeled;
\item Start-up time is closely captured;
\item Liquid plug velocities, length are qualitatively predicted by simulation;
\item In agreement with the simulations, the spatial variation of the liquid temperature appears to be generally smooth, with sharp boundary layers near the plug ends;
\item A long-term functioning prediction for \SI{185}{W} evaporator power is reported. The steady state is established just after \SI{1}{min}. The continuous oscillations without stopovers are observed, which is a highly efficient PHP regime.
\end{itemize}

A complete experimental validation on the steady state is left for future, when the data obtained in the ISS experiments will be available.

\section*{Acknowledgments}
The present work is carried forward in the framework of the European Space Agency Microgravity Application Programme Project entitled Two-phase passive thermal devices for deployable Space Systems (TOPDESS, Project number 4000128640). The authors would like to thank all the members of the International Scientific Team on PHP led by Prof. M. Marengo and all the scientists and engineers who are contributing to the development of the Heat Transfer Host experiment on PHPs. VN acknowledges the financial support from CNES granted in the framework of the GDR MFA.
\appendix
\section{PHP model equations}\label{AppFEC}

The FEC model \cite{JHT11,IarATE17,IHPC18VN,MST19,ATE21} implemented in the present version of the CASCO software uses the following main assumptions:
\begin{enumerate}
\item The two-phase flow regime is the plug-slug flow. Vapor and film (if any) occupy all the tube cross-section in the vapor domains called vapor bubbles; the incompressible liquid occupies all the tube within the liquid domains called liquid plugs.\label{ass0}
\item The liquid films are of the constant thickness $\delta_f$. Each film is deposited by the receding meniscus and absorbed by the advancing meniscus. The films may be partially or completely evaporated (except in the condenser, see assumption \ref{assCond}). Arbitrary number of dry spots per bubble are allowed. The film mass exchange is controlled by the interfacial energy balance. Film evaporation leads to a receding of the film edge until it retracts from the superheated area and film condensation causes its advance. If evaporation occurs from the film part adjacent to its edge but condensation occurs on another part of the film, the condensation does not impact the edge dynamics; the condensed mass is equally shared between the neighboring liquid plugs for mass conservation.\label{assFilms}
\item In addition to the film mass exchange, there is a mass exchange from each meniscus proportional to the local superheating at the meniscus position. It plays an important role when the films are absent. \label{assMen}
\item If under-saturated (i.e. $p_i<p_{sat}(T_i)$), the vapor of the bubble $i$ obeys the ideal gas equation of state. When $p_i$ reaches $p_{sat}(T_i)$, the vapor stays in the saturated state as long as its pressure change calculated from the saturated curve stays smaller than that calculated with the ideal gas EOS \cite{IHPC18VN}. \label{assState}
\item Bubbles disappear when their length reaches a (small) threshold values, $L_v^{thr}$. Similarly, a plug deletion threshold length $L_l^{thr}$ is introduced.\label{assDel}
\item Bubble of the length $L_{nucl}$ is generated at any point inside the plug if the local superheating exceeds a nucleation barrier $\Delta T_{nucl}$. The bubbles cannot however be generated very close to the plug menisci, at a distance smaller than $L_{nucl,min}$. The pressure used for the local superheating calculation varies linearly along the plug. \label{assAdd}
\item In the present CASCO version, the temperature of the internal tube walls in the condenser section is assumed to be fixed,  $T_w=T_c$ (i.e. the cooler is ideally efficient). Consequently, the films always surround the bubbles in the condenser. \label{assCond}
\end{enumerate}

CASCO has a capability to simulate the multi-source heaters and coolers. The respective numbers are denoted $N_e$ for hot and $N_c$ for cold sources. Since there is always an adiabatic zone between any hot and cold zones, the adiabatic zones number is $N_a=N_e+N_c$. The closed PHP channel is opened and projected to the straight $x$-axis. The evaporator, condenser, and adiabatic sections follow each other sequentially in a periodical way along $x$. $x=0$ point corresponds to the beginning of the first evaporator section in the first period (Fig. \ref{Init}). The total period number is $N_{turn}$. The total PHP channel length is thus $L_t=N_{turn}L_p+L_{fb}$, where $L_p$ is a period
length and $L_{fb}$ is that of the feedback section that closes the PHP loop.
The period length is
\begin{equation}\label{Lpa}
  L_p=\sum\limits_{k=1}^{N_e}L_e^k+\sum\limits_{k=1}^{N_c}L_c^k+\sum\limits_{k=1}^{N_a}L_a^k.
\end{equation}
where $L_e^k$, $L_c^k$, $L_a^k$ are the lengths of the respective zones. To complete the PHP geometry, one needs to define $X_{fb}$, a distance from the beginning of the first evaporator of the last period to the beginning of the feedback.

Contrary to earlier approaches, the menisci can freely displace along it so their coordinates can be even negative. It is assumed that a bubble and a plug to the right of it have the same number $i$. The right end of the $i$-th plug (i.e. the left end of a bubble next to it) is $X_{i,{next}}^l=X_{i+1}^l $ for $i<M$, where $M$ is the total number of bubbles. The closed loop PHP is implemented with the periodicity condition $X_{M,{next}}^l=X_1^l+L_{t}$. To find a physical position inside the PHP channel, one needs to reduce its abscissa by finding a remainder of its division by $L_{t}$.

To complete the geometry description, one needs to fix one more independent parameter: the feedback offset $X_{fb}$ with respect to the beginning of the last period (Fig.~\ref{Init}). It also fixes the position of the first evaporator inside a period.

The velocity $V_i$ of the center of mass of the $i$-th plug is determined from its momentum balance
\begin{align}
&\frac{\mathrm{d}}{\mathrm{d}t}(m_{{l},i}V_i)=(p_i-p_{i,{next}})S - F_i,\label{dVi}
\end{align}
where $S=\pi r^2$ and the viscous friction $F_i$ is given by Poiseuille expression for small $Re=2V_ir/\nu$ or by the Blasius correlation \cite{shafii1} for the turbulent flow:
\begin{align}
F_i&=\frac{\pi r V_i^2m_{l,i}}{S}\left\{ \begin{array}{ll}
16/Re, & 0<Re<1180\\
0.079\,Re^{-0.25}, & Re \geq 1180
\end{array} \right.\label{F}
\end{align}
The plug mass varies in time. This impacts the velocities of left $\dot X_{i}^l$ and right $\dot X_{i}^r$ bubble menisci that should thus be determined from the set of equations
\begin{align}
  V_i&=\frac{1}{2}\left(\dot X_i^r+\dot X_{i,next}^l\right),\label{ViFEC}\\
  \dot m_{l,i}&=\rho_l S \left(\dot X_{i,next}^l-\dot X_{i}^r\right).\label{ml}
\end{align}
where the plug mass change rate $\dot m_{l,i}$ is defined below in \eqref{dml} and the dot means the time derivative.

As the bubbles can extend over several evaporator sections of the channel, several dried areas in the liquid film (``holes'') per bubble and, accordingly, several film pairs can be formed. Their left and right edge positions are denoted $X_{f,i}^{l,k}$ and $X_{f,i}^{r,k}$ for the $k$-th hole. The effective evaporator is a part of the wall inside the bubble $i$, which is either dry or along which $T_w$ exceeds $T_{sat}(p_i)$. The number of effective evaporators is denoted $N_{e,i}$ and their borders as $X_{e,i}^{s,k}$ ($s=l,r$). Note that the number of effective condensers is $N_{e,i}+1$.

The vapor mass change of the $i$-th bubble caused by the evaporation of right and left films of the $k$-th effective evaporator are defined by the local interfacial energy balance (assumption \ref{assFilms}):
\begin{align}
\dot{m}_{{f,e},i}^{{r},k}&=\frac{U_{f}\pi r_f}{{\cal L}}\int\limits_{X_{{f},i}^{{r},k}}^{X_{{e},i}^{{r},k}}[T_{w}(x)-T_{sat}(p_i)]\mathrm{d}x,\label{eq:mf_er}\\
\dot{m}_{{f,e},i}^{{l},k}&=\dfrac{U_{f}\pi r_f}{{\cal L}}\int\limits_{X_{{e},i}^{{l},k}}^{X_{{f},i}^{{l},k}}[T_{w}(x)-T_{sat}(p_i)]\mathrm{d}x.\label{eq:mf_el}
\end{align}
where $r_f=r-\delta_f$ and $U_f=\varphi\lambda_l/\delta_f$, where $\varphi\simeq 0.47$ is the film form factor \cite{IJHMT10}. The vapor mass change caused by the condensation in the $k$-th effective condenser is
\begin{equation} \label{eq:mf_c}
\dot{m}_{{f,c},i}^{k}=\dfrac{U_{f}\pi r_f}{{\cal L}}
\begin{cases}
\int\limits_{X_{i}^{{l}}}^{X_{{e},i}^{{l},k}}[T_{w}(x)-T_{sat}(p_i)]\mathrm{d}x,&\text{if }k=1\\
\int\limits_{X_{{e},i}^{{r},k-1}}^{X_{{e},i}^{{l},k}}[T_{w}(x)-T_{sat}(p_i)]\mathrm{d}x,&\text{if }1<k \leq N_{{e},i}\\
\int\limits_{X_{{e},i}^{{r},k-1}}^{X_{i}^{{r}}}[T_{w}(x)-T_{sat}(p_i)]\mathrm{d}x,&\text{if }k=N_{{e},i}+1
\end{cases}
\end{equation}
The vapor mass change caused by the $s$-th ($s=l,r$) meniscus phase change (assumption \ref{assMen}) is
\begin{equation} \label{mm_i}
\dot{m}_{{m},i}^s=\dfrac{U_{m}\pi r L_{m}}{{\cal L}}[ T_{w}(X_i^s)-T_{sat}(p_i) ],
\end{equation}

To implement the assumption \ref{assFilms}, the film dynamics is described as
\begin{align}
\dot{X}_{{f},i}^{{l},k}&=
\begin{cases}
\dot{X}_{i}^{{l}}, &\textrm{ if }X_{f,i}^{l,k} = X_{i}^{l} \textrm{ and } \dot{X}_{i}^{l} \geq 0\\
\dot{X}_{i}^{{r}}, &\textrm{ if } X_{f,i}^{l,k} = X_{i}^{r} \textrm{ and } \dot{X}_{i}^{r} < -\dfrac{\dot{m}_{f,e,i}^{l,k}}{\rho S_f},\\
-\dfrac{\dot{m}_{{f,c},i}^{k}}{2\rho_l S_f}, &\textrm{ if } X_{{e},i}^{{l},k} = X_{{f},i}^{{l},k} \textrm{ and } X_{{f},i}^{{l},k} < X_{{f},i}^{{r},k}\\
-\dfrac{\dot{m}_{{f,}e,i}^{{l},k}}{\rho_lS_f}, &\textrm{ otherwise}
\end{cases} \label{eq:xfl}\\
\dot{X}_{{f},i}^{{r},k}&=
\begin{cases}
\dot{X}_{i}^{{r}}, &\textrm{ if }X_{f,i}^{r,k} = X_{i}^{r} \textrm{ and } \dot{X}_{i}^{r} \leq 0\\
\dot{X}_{i}^{{l}}, &X_{f,i}^{r,k} = X_{i}^l \textrm{ and } \dot{X}_{i}^{l} > \dfrac{\dot{m}_{f,e,i}^{r,k}}{\rho S_f},\\
\dfrac{\dot{m}_{{f,c},i}^{k+1}}{2\rho_lS_f}, &\textrm{ if }X_{{e},i}^{{r},k} = X_{{f},i}^{{r},k}  \textrm{ and } X_{{f},i}^{{l},k} < X_{{f},i}^{{r},k}\\
\dfrac{\dot{m}_{{f,e},i}^{{r},k}}{\rho_lS_f}, &\textrm{ otherwise}
\end{cases}\label{eq:xfr}
\end{align}
where $S_f=\pi (r^2-r_f^2)$ is the film cross-section area, and the order of lines is meaningful (2nd option holds if the 1st is invalid, the 3rd if first two are invalid, etc.).

The vapor description (assumption \ref{assState}) was introduced by \citet{IHPC18VN}. The bubble volume can be determined as
\begin{equation}\label{Omega}
\Omega_i=(S-S_f)({X}_i^r-{X}_i^l)+S_f\sum\limits_{k=1}^{N_{e,i}}\left( X_{f,i}^{r,k}- X_{f,i}^{l,k}\right).
\end{equation}
If the vapor is in the under-saturated state (called also superheated, $p_i<p_{sat}(T_i)$), it obeys the ideal gas EOS
\begin{subequations}\label{eos}
\begin{equation}\label{eosi}
  p_i=m_iR_vT_i/\Omega_i.
\end{equation}
The total mass change rate for the $i$-th bubble in this regime is
\begin{equation} \label{mit}
\dot{m}_i=\dot{m}_{{m},i}^l+\dot{m}_{{m},i}^r+\dot{m}_{{f,c},i}^{N_{{e},i}+1}+\sum^{N_{{e},i}}_{k=1} \left(\dot{m}_{{f,e},i}^{{l},k}+\dot{m}_{{f,e},i}^{{r},k}+\dot{m}_{{f,c},i}^{k}\right).
\end{equation}

Once $p_i$ calculated with \eqref{eosi} rises above $p_{sat}(T_i)$, the saturated state is assumed to be attained so
\begin{equation}\label{psat}
p_i=p_{sat}(T_i)
\end{equation}
is assumed. If the bubble remains in the saturated state, its mass change rate is calculated as
\begin{equation}\label{mvsat}
\dot{m}_i=\dot\Omega_i\frac{ p_{sat}(T_i)}{R_vT_i}.
\end{equation}
\end{subequations}

The vapor energy balance depends on its thermodynamic state. When it is under-saturated, the energy equation for the $i$-th bubble is \cite{shafii1}
\begin{subequations}\label{en}
\begin{equation}\label{ensup}
m_ic_{v,v}\dot{T}_i=\dot m_iR_vT_i+P^{sens}_i-p_i\dot\Omega_i,
\end{equation}
where
\begin{equation}\label{Qsens}
P^{sens}_i=2\pi rU_v\sum\limits_{k=1}^{N_{e,i}}\int\limits_{X_{f,i}^{l,k}}^{X_{f,i}^{r,k}}[T_w(x)-T_i]\text{d}x,
\end{equation}
with $U_v=Nu_v\lambda_v/(2r)$ and $Nu_v=6$ \cite{Gully14}.

When the vapor is at saturation, one can assume that
\begin{equation}\label{ensat}
\dot{T}_i=0,
\end{equation}
\end{subequations}
and the vapor pressure does not change either.

An additional criterion is needed to let the vapor leave the saturation state by comparing the pressure derivatives obtained for under-saturated and saturated states by using the same temperature change $\dot{T}_i$ (the one given by Eq.~\eqref{ensup} is taken). First, consider the pressure derivative $\dot p_v$ for the under-saturated state. It is obtained from Eq.~\eqref{ensup}. This equation can be reduced by using both EOS \eqref{eosi} and Mayer's relation  $c_{v,p}=c_{v,v}+R_{v}$ valid for the ideal gas:
\begin{equation}\label{dotp}
 \dot p_v=\frac{p_i}{T_i} \frac{\gamma}{\gamma-1}\dot T_i-\frac{P^{sens}_i}{\Omega_i}.
\end{equation}
The second pressure variation
\begin{equation}\label{dpsat}
\dot p_{sat}=\left.\frac{\textrm{d}p}{\textrm{d}T}\right|_{sat}\dot T_i,
\end{equation}
is for the saturated state. The vapor stays at saturation while $\dot p_v\geq\dot p_{sat}$ and leaves it when $\dot p_v<\dot p_{sat}$.

As specified in the assumption \ref{assFilms}, some of the liquid condensed to the film does not serve to increase its mass and should be thus shared between the neighboring plugs. The clauses of the equation below conform to those of Eqs. (\ref{eq:xfl}, \ref{eq:xfr}) written for the given $k$:
\begin{multline}\label{consec}
2\dot{m}_{cons,i}=\frac{1}{2}\left(\dot{m}_{f,c,i}^{1}+
\dot{m}_{f,c,i}^{N_{e,i}+1}\right)\\+
\frac{1}{2}\sum\limits_{k=1}^{N_{e,i}}\left[
\left\{\begin{array}{ll}
\dot{m}_{f,c,i}^{k}, &\textrm{ if 2nd or 4th lines in \eqref{eq:xfl}} \\
0, &\textrm{ if 1st or 3rd lines in \eqref{eq:xfl}}\end{array}\right\}\right.\\+\left.
\left\{\begin{array}{ll}
\dot{m}_{f,c,i}^{k+1}, &\textrm{ if 2nd or 4th lines in \eqref{eq:xfr}}\\
0, &\textrm{ if 1st or 3rd lines in \eqref{eq:xfr}}\end{array}\right\}\right].
\end{multline}
Finally, one introduces the modified $s$-th ($s=l,r$) meniscus evaporation rate accounting for the above terms,
\begin{equation}\label{mm_imod}
\dot m_{m,i}^{s,*}=\dot m_{m,i}^{s}+\dot{m}_{cons,i}.
\end{equation}

The plug mass change rate (assumptions \ref{assFilms},\ref{assMen}) is then
\begin{align}
&\dot m_{l,i}=-\dot m_{m,i,\,next}^{l,*}-\dot m_{m,i}^{r,*}\nonumber\\&-\rho_lS_f\Bigg[\left\{ \begin{array}{ll}
0, & \mbox{if }\dot{X}_i^r<0\mbox{ and no film at }X^r_i,\\
\dot{X}_i^r,  & \mbox{otherwise},\end{array}\right\}\nonumber\\& -\left\{
\begin{array}{ll}
0, & \mbox{if }\dot{X}_{i,\,next}^l>0\mbox{ and no film at }X^l_{i,\,next},\\
\dot{X}_{i,\,next}^l, & \mbox{otherwise},
\end{array} \right\}\Bigg].\label{dml}
\end{align}

The wall temperature $T_w$ is determined \cite{IJHMT16} from the 1D heat diffusion equation
\begin{equation}\label{Ts}
\frac{\partial T_w}{\partial t}=D_w\frac{\partial^2 T_w}{\partial x^2}+\frac{j_w}{\rho_w c_w}
\end{equation}
solved within the evaporator and adiabatic section;
\begin{equation}\label{js}
j_w =  \frac{2\pi}{S_w}\begin{cases}
r_e\, q_{s}(x)- r\,q_{fluid}(x), &\mbox{ if } x\in\mbox{evaporator},\\
-r\,q_{fluid}(x), &\mbox{ if } x\in\mbox{adiab. sec.}
\end{cases} \end{equation}
is the equivalent volume heat flux, where $q_s$ is the heat flux from the evaporator spreader to the tube of the external radius $r_e$; $S_w=\pi(r_e^2-r^2)$.

The heat flux
\begin{equation}\label{qfluid}
q_{fluid}(x)=U_{fluid}(x)[T_w(x)-T_{fluid}(x)]+q_m(x)
\end{equation}
is transferred from the internal tube wall to the fluid, where
\begin{equation}\label{Tfluid}
T_{fluid}(x) =  \begin{cases}
T_i, &\mbox{ if } x\in\mbox{dry area of bubble }i,\\
T_{sat}(p_i), &\mbox{ if } x\in\mbox{film }i,\\
T_l(x), &\mbox{ if } x\in\mbox{liquid}.
\end{cases} \end{equation}
The heat exchange coefficient $U_{fluid}$ is either $U_v$, $U_f$ or $U_l$ for the respective regions; $U_l=Nu_l\lambda_l/(2r)$, where $Nu_l$ is given by the \citet{Gnielinski76} correlation. The flux is not injected into the vapor ($U_v\to 0$) while it remains at saturation in agreement with \eqref{ensat}. The heat flux corresponding to evaporation at each meniscus $\dot m_{m,i}^{s}$ is injected at the meniscus location,
\begin{equation}\label{qm}
q_m(x)=U_m L_m\sum_{i=0}^M\sum_{s=l,r}\left[T_w(x)-T_{sat}(p_i) \right]\delta\left(x-X_i^s\right),
\end{equation}
where $\delta(x)$ is the Dirac delta function. Based on the theory of contact line evaporation \cite{ATE21}, it is assumed that $U_m\simeq 0.3U_f$ and $L_m\simeq\SI{0.2}{mm}$.

The temperature distribution in the liquid plug $T_{l,i}=T_{l,i}(x,t)$ where $x\in\left(X_i^r,X_{i,next}^l\right)$ is governed by the heat diffusion equation \cite{shafii1} where the convective heat exchange with the wall is included:
\begin{equation}\label{hd}
\frac{\partial T_{l,i}}{\partial t}=D_l \frac{\partial^2 T_{l,i}}{\partial x^2}+D_l\frac{Nu_l}{r^2}(T_{w}-T_{l,i}).
\end{equation}
The boundary conditions for Eq.~(\ref{hd}) are given at the menisci,
\begin{equation}\label{bc} \begin{split}
  T_{l,i}(x=X_i^r)&=T_{sat}(p_i), \\
T_{l,i}(x=X_{i,next}^l)&=T_{sat}(p_{i,next}).
\end{split}\end{equation}

Different thermal models of evaporator can be used. Here, it is used a massive highly conductive (and thus isothermal) evaporator block (spreader) of the thermal mass $C_s$ incorporating the heating elements. There is a contact thermal resistance between the tubes and the massive evaporator. The corresponding thermal conductance per area is $U_{s}$ so the heat flux from the spreader to the tube is
\begin{equation}\label{qse}
 q_{s}(x)=U_{s}[T_s-T_w(x)].
\end{equation}
The spreader temperature $T_s$ obeys the energy balance
\begin{equation}\label{Tseq}
C_s\frac{\mathrm{d}T_s}{\mathrm{d}t}=P_e-2\pi r_e\int_0^{L_t}\left.\begin{cases}q_{s}(x),
 &\mbox{ if } x\in\mbox{evaporator},\\
0, &\mbox{ otherwise},
\end{cases}\right\}\mathrm{d}x.
\end{equation}


\end{document}